\documentclass[twocolumn]{aastex701}

\usepackage{multirow}
\usepackage{multirow, physics, float, amsmath, mathtools}

\usepackage{graphicx, placeins}
\usepackage{hyperref}

\newcommand{\lsuperscript}[2]{\textnormal{{\textsuperscript{#1}}#2}}
\newcommand{\differentiald}{\text{d}}

\accepted{September 29, 2025}

\begin{document}
\title{An excess of luminous white dwarfs in the peculiar Galactic globular cluster NGC~2808}

\author[orcid=0009-0007-9782-8777]{Laksh Gupta}
\affiliation{School of Arts $\&$ Sciences, Ahmedabad University, Ahmedabad - 380009, India.}
\email[show]{laksh.g@ahduni.edu.in}  

\author[orcid=0000-0001-8182-9790]{Samyaday Choudhury}
\affiliation{School of Arts $\&$ Sciences, Ahmedabad University, Ahmedabad - 380009, India.}
\affiliation{International Centre for Space and Cosmology, Ahmedabad University, Ahmedabad - 380009, India.}
\email[show]{samyaday.choudhury@ahduni.edu.in}

\author[orcid=0000-0002-0882-7702]{Annalisa Calamida}
\affiliation{Space Telescope Science Institute, 3700 San Martin Drive, Baltimore, MD 21218, USA.}
\affiliation{INAF - Osservatorio Astronomico Capodimonte, Salita Moiariello, 16, 80131 Napoli, Italy.}
\email{calamida@stsci.edu}

\author[orcid=0000-0002-8878-3315]{Christian I. Johnson}
\affiliation{Space Telescope Science Institute, 3700 San Martin Drive, Baltimore, MD 21218, USA.}
\email{chjohnson1@stsci.edu}

\author[orcid=0000-0003-1149-3659]{Domenico Nardiello}
\affiliation{Dipartimento di Fisica e Astronomia ``Galileo Galilei'' -- Universit\`a degli Studi di Padova, Vicolo dell'Osservatorio 3, I-35122 Padova, Italy}
\affiliation{INAF -- Osservatorio Astronomico di Padova, Vicolo dell'Osservatorio 5, Padova I-35122, Italy}
\email{domenico.nardiello@unipd.it}

\begin{abstract}

We study the white dwarf (WD) cooling sequence of the Galactic Globular Cluster (GGC) NGC~2808 by using deep near-UV data from the Hubble Space Telescope and theoretical models, to investigate if this cluster hosts an excess of WDs. Excess in WDs is a rare phenomenon that has been found to exist only in a few GGCs. We compared star counts from different evolutionary phases on the near-UV color-magnitude diagram to evolutionary times predicted by BaSTI models. The investigation was carried out over a region within a radii of 1.5 $\arcmin$ of the cluster center and a region of similar dimension located 5$\arcmin$ away. We find a WD excess of $\approx$ 60 - 70\% when comparing star counts and evolutionary models of the WD cooling sequence to the main-sequence turn-off, and by using different values and fractions of Helium enhancement. This excess decreases to $\approx$ 30 - 40\% when the WD cooling sequence is compared to the horizontal branch. The WD excess is slightly larger in the internal field that covers the cluster center; however, the difference with the external field is compatible within the uncertainties. We argue that this excess is possibly related to the existence of SCWDs and Helium-core WDs in NGC~2808, and might be directly associated to the extended blue horizontal branch of this GGC.


\end{abstract}



\section{Introduction}

\begin{figure*}
    \centering
    \includegraphics[width=0.7\linewidth]{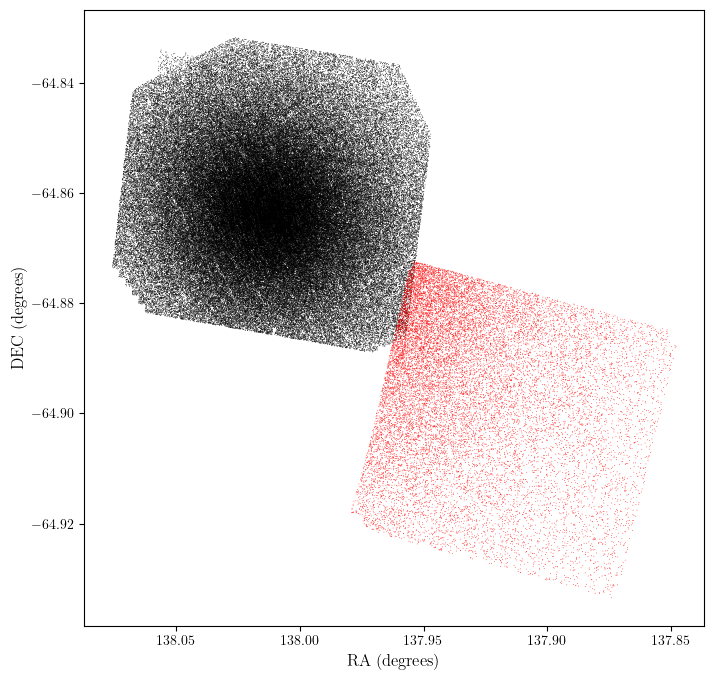}
    \caption{Stars from the HST HUGS catalog for NGC~2808 are plotted in the RA and DEC plane as black dots ('internal field'), while red dots indicate stars detected and measured from HST program GO-15857 (`external field'). The corners of the internal field are cut since only the region covered by photometry in all WFC3-UVIS near-UV filters is represented.}
    \label{SDPNGC2808}
\end{figure*}

\begin{figure*}
    \centering
    \includegraphics[width=0.7\linewidth]{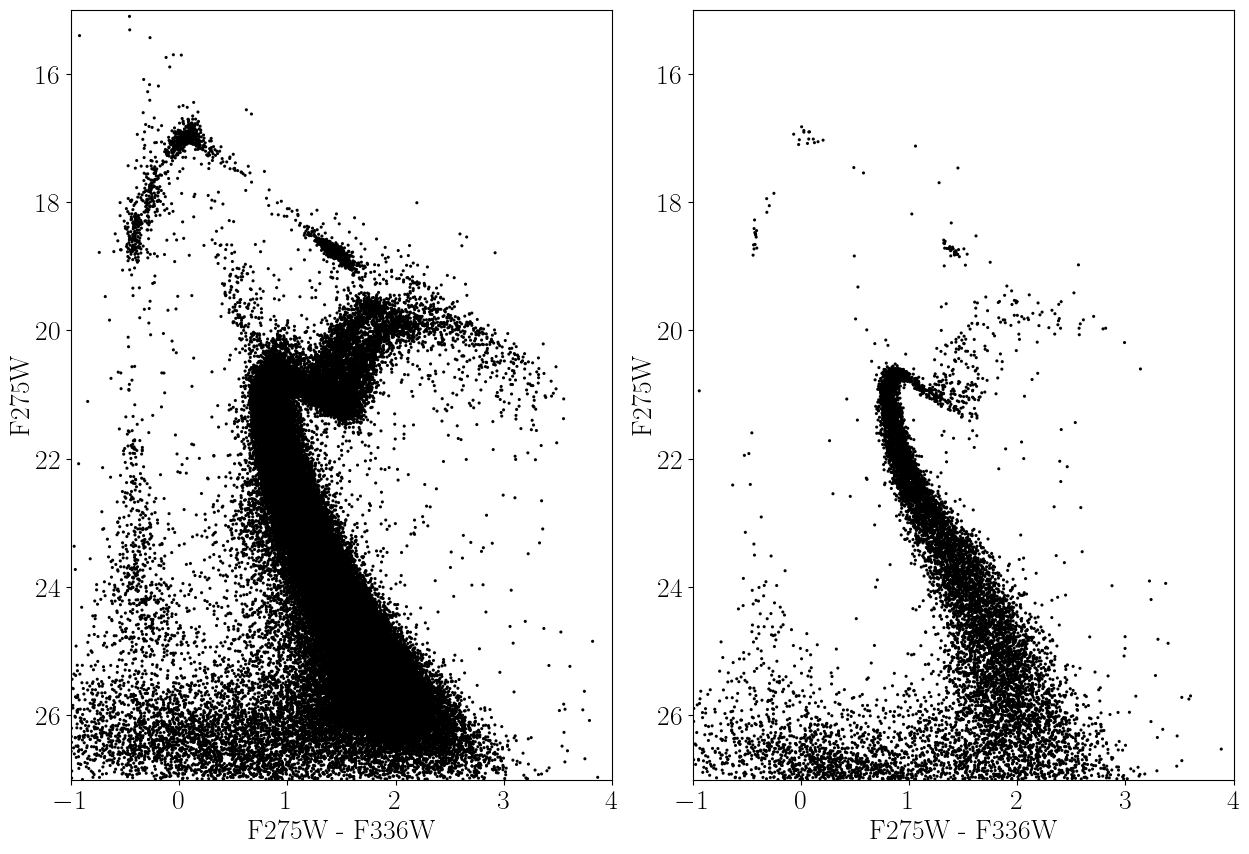}
    \caption{Left -- $F275W,\ F275W - F336W$ CMD for the internal field with 145,610 objects whose error is $<$ 1~mag in $F275W$ and $F336W$. Right -- Same plot as the left for the external field with 10,897 objects.
    }
    \label{336CMD Intro}
\end{figure*}


White Dwarf (WD) stars are hot (5000 - 80000 K) and dense ($10^9$ kg/m$^3$) stellar remnants. They represent the final evolutionary stage of low-mass (M $\lesssim$ 8 M$_{\odot}$) stars, and retain the imprints of their progenitors' evolutionary histories (\citealt{HansenLiebert03}). The absence of energy generation aids in describing their evolution as a simple cooling process. For this reason, the time a WD takes to cool (the cooling time) is { inversely} proportional to its luminosity. Consequently, WDs are excellent cosmic chronometers in various stellar populations such as the Galactic disk and star clusters \citep{tremblay2014}. The formation and cooling of WDs - whether with Carbon-Oxygen or Helium core - are closely linked to post main-sequence processes. These include factors such as binarity, mass-loss in red giant or asymptotic giant phase, and variations in Helium abundance. 
{ WDs are fainter than most other stars in a globular cluster, and in order to identify them and understand their evolutionary pathways in these systems, deep photometric data and high spatial resolution are a necessity.}

WDs have been extensively observed through imaging and characterized through spectroscopy in the Galactic disk \citep{eisenstein06, kepler07, koester09}. 
Disk WDs presented by \cite{Kleinman2013} have a mean mass of $\sim$ 0.6 M$_{\odot}$ (DA: Hydrogen envelope) and $\sim$ 0.68 M$_{\odot}$ (DB: Helium envelope); however, the disk WD mass distribution showed a secondary peak at $\approx$ 0.4 M$_{\odot}$ \citep{kepler07}, with $\approx$ 4\% of WDs being less massive than this value \citep{rebassa11}. These low-mass WDs probably have Helium-cores and, given that their ages must be less than a Hubble time, they should result from close-binary interactions after a post-common envelope phase or from the merging of two very low-mass Helium-core WDs (He-core WDs) \citep{han08}. The binary scenario is supported by the finding of numerous low-mass WDs in binary systems in the disk, with another WD, a neutron star, or a subdwarf B star as a companion \citep{marsh95, maxted02}. Single He-core WDs have also been observed \citep{marsh95, kilic07}, while extremely low-mass He-core WDs {(M$_{\ast}\approx$ 0.2 M$_{\odot}$)} have only been found in binary systems \citep{kilic12}. 

Photometric evidence of He-core WDs was found in the bulge \citep{Calamida2014} and in a few Galactic Globular Clusters (GGCs), such as $\omega$ Cen \citep{2005Monelli, cal08mainpaper, 2013Bellini, scalco2024}, NGC~6752 \citep{ferraro03}, and NGC~6397 \citep{Strickler_2009}, and in the old metal-rich open cluster NGC~6791 \citep{kalirai07, Bedin2008}. At the same time, a population of extreme horizontal branch (EHB) stars was observed in most of these clusters, also in a metal-rich environment as NGC~6791, and a few EHBs were identified in the bulge by \citet{zoccali03}, \citet{busso05}, and \citet{Calamida2013}. 

Studies using deep Hubble Space Telescope (HST) imaging and updated WD models \citep{2015Althaus} have proposed the presence of slowly cooling WD (SCWDs) in a few GGCs, such as M~13 \citep{2021Chen} and NGC~6752 \citep{2022Chen}. However, no evidence of SCWDs was found in M~3, which has very similar age and metallicity compared to M~13, but no EHB. These SCWDs are thought to be the aftermath of blue HB or EHB stars that evolve as AGB manqu\'e { (see \citealt{1993Dorman} and Fig.~6 of \citealt{2021Chen})}, or post-early AGB stars and skip the third dredge-up; in this way they retain a residual hydrogen envelope thick enough for thermonuclear burning when they reach the WD stage, resulting in a slower cooling along the sequence \citep{2015Althaus}. Hence, for GGCs that host them, the brighter WD sequence would have an excess of WDs. From the aforementioned studies, a connection seems to be emerging between the presence of blue HB or EHB stars and that of He-core or SCWDs. 

In GGCs, WDs provide statistically significant samples to constrain the validity of the physical assumptions incorporated to construct WD cooling models compared with open star clusters. Additionally, GGCs are situated at a well-constrained distance, are relatively homogeneous populations (in terms of age and metallicity spread) compared with the Galactic disk or bulge, rendering them excellent astrophysical laboratories to study WDs. In this study, we use deep near-UV HST data to analyze the WD cooling sequence in the highly unusual cluster NGC~2808, which has one of the most populated and extended HB sequence among GGCs \citep{bedin2000}.

\emph{HST} photometric investigations revealed that NGC~2808 main-sequence (MS) splits into a blue, an intermediate, and a red sequence \citep{Piotto2007}. Additionally, NGC~2808 shows an extended HB with distinct components: a red HB (RHB) and a blue tail divided into three groups \citep{sosin1997, bedin2000}. Therefore, it was suggested that NGC~2808 experienced multiple episodes of star formation, each with varying levels of helium enrichment and with the bluest MS representing the most enhanced population \citep{Piotto2007, Milone2012, Milone2015quintuple}. Helium enrichment could also explain the observed morphology of HB \citep{dantona04,dantona05, Lee05, Prabhu_2021}. Very recently, a small spread of metallicity in the cluster was reported \citep[see][]{2023Lardo, 2025Latour}. Table~\ref{Table NGC~2808 Parameters} summarizes the basic properties of NGC~2808.

To investigate if NGC~2808 hosts He-core WDs or SCWDs, we use star counts and compare them to evolutionary time following the method presented in \citet{2007Castellani} and  \citet{cal08mainpaper}. Ideally, the ratio of star counts to evolutionary times of two different evolutionary stages must be of the same order of magnitude. A deviation from this trend would suggest an excess or deficiency of stars observed. 

To count stars in various evolutionary phases, such as Main Sequence Turnoff (MSTO), Horizontal Branch (HB), and WD cooling sequence, we use the ${F275W}$ vs. $F275W$ - $F336W$ CMDs. We select ${F275W}$ and ${{F336W}}$ photometry since HB stars and WDs are hot and so brighter in these blue filters. 

The organization of the paper is as follows. Section~\ref{section:observations and data} describes the data reduction, calibration and the artificial star tests. The method and analysis are presented in Section~\ref{section:analysis}, along with evolutionary time analysis of the MSTO, HB, and WD stars corresponding to canonical models. Evolutionary time analysis corresponding to Helium-enhanced models are presented in Section~\ref{He-enhanced Crossing Times for HB and MSTO stars}, followed by star count-evolutionary time calculations in Section~\ref{SC-CT}. The discussion is presented in  Section~\ref{section:discussion} and conclusions in Section~\ref{section:summary}. 

\section{Observations \& Data Reduction}\label{section:observations and data}


\begin{deluxetable}{ccc}
\setlength{\tabcolsep}{15pt}
\tablecaption{{Table shows the positional, photometric, and structural parameters of NGC~2808.}}
\tablehead{ \colhead{Parameter} & \colhead{Value} & \colhead{References} }
\startdata
$\alpha$ ({ deg,} J2000) & 138.0071 & 1 \\
$\delta$ ({ deg,} J2000) & -64.8645 & 1 \\
$\mu_\alpha$ { (mas/yr$^{-1}$)} & $0.994 \pm 0.024$ & 2 \\
$\mu_\delta$ { (mas/yr$^{-1}$)} & $0.273 \pm 0.024$ & 2 \\
$M_V$ (mag)$^{\text{a}}$ & -9.4 & 3 \\
$\mu_0$ (mag)$^{\text{g}}$ & 15.05 & 3 \\
$[\text{Fe}/\text{H}]$ (dex) & -1.14 & 3 \\ 
$\log(t_h)^{\text{e}}$ & 8.24 & 3 \\
$r_c$ (arcmin)$^{\text{b}}$ & 0.26 & 4 \\
$r_h$ (arcmin)$^{\text{c}}$ & 0.86 & 5 \\
$r_t$ (arcmin)$^{\text{d}}$ & 21.97 & 5 \\
$E(B-V)^{\text{f}}$ & $0.19 \pm 0.03$ & 6 \\
Age (Gyr) & $10.9 \pm 0.7$ & 7 \\ 
\enddata
\tablecomments{\lsuperscript{a}{Total visual magnitude}, \lsuperscript{b}{Core radius}, \lsuperscript{c}{Half-mass radius}, \lsuperscript{d}{Tidal radius}, \lsuperscript{e}{Log of relaxation time}, \lsuperscript{f}{Reddening}, \lsuperscript{g}{True distance modulus}. \\ References: (1) \citet{Gaia2018}, (2) \citet{Vasiliev}, (3) \citet{Harris2010}, (4) \citet{Trager}, (5) \citet{McLaughlin}, (6) \citet{SchlafyFinkbeiner}, (7) \citet{2016Massari}}
\label{Table NGC~2808 Parameters}
\end{deluxetable}

\begin{figure*}
    \centering
    \includegraphics[width=0.7\linewidth]{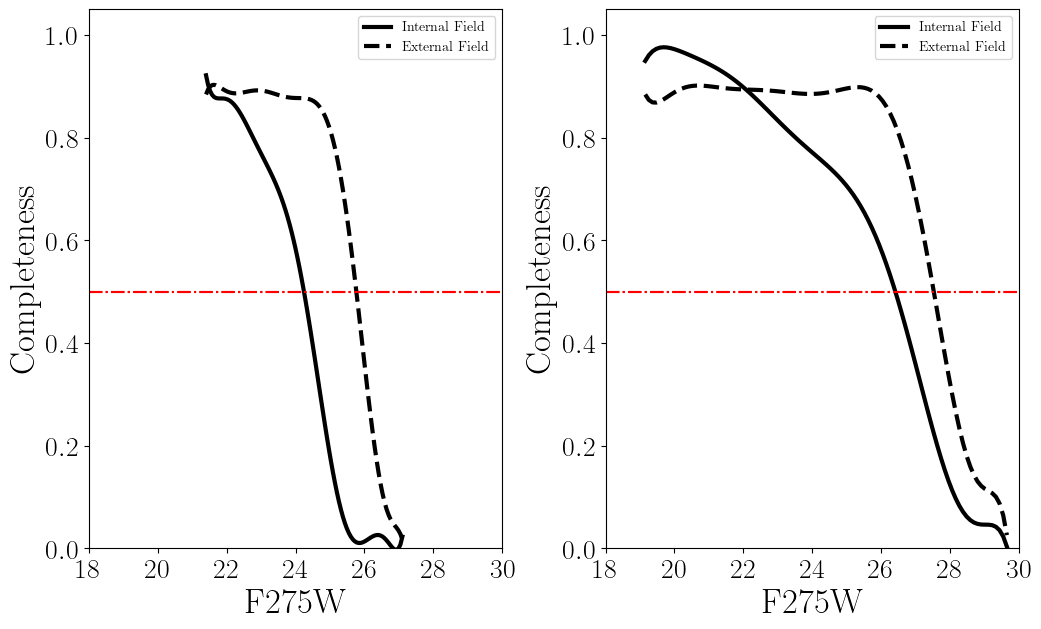}
    \caption{Left -- The black solid and dashed lines indicate the completeness curves for WDs observed in the $F275W$ filter for the internal and the external field, respectively. The dashed-dotted horizontal line indicates a completeness level of 50\%. Right -- Same but for the completeness of the other evolutionary phases, such as MSTO, RGB and HB.}
    \label{AST Curves}
\end{figure*}




We downloaded HST data for NGC~2808 
from the ``Hubble Space Telescope Ultraviolet Legacy Survey of Galactic Globular Clusters'' (HUGS) database { \citep{https://doi.org/10.17909/t9810f, Piotto2015, Nardiello2018}}. The catalog includes photometry in five filters: WFC3-UVIS $F27W5$, $F336W$, and $F438W$ (program GO -2605, PI: Piotto) and ACS-WFC $F606W$, and $F814W$ (program GO-10775, PI: Sarajedini)\footnote{The HST data can be obtained from the Milkulski Space Archive for Telescopes at the Space Telescope Science Institute via \href{https://archive.stsci.edu/prepds/hugs/}{https://archive.stsci.edu/prepds/hugs/}}. The catalog provides photometric errors (labeled as $d$ parameter), a sharpness parameter ($s$), and the goodness of fit of the source with the HST effective Point Spread Function (ePSF, $q$ parameter). It also provides the cluster membership probability ($P_{\mu}$) of each object based on the measured proper motions (PMs). These data cover an area with $r\lesssim$ 1.5$\arcmin$across the cluster center, and we label it as the 'internal field' (see Fig.~\ref{SDPNGC2808}).


We also downloaded and reduced WFC3-UVIS images from program GO-15857 (PI: Bellini) for a field at $\approx$ 5' from the cluster center. 
{ This is labeled as the `external field' (see Fig.~\ref{SDPNGC2808}); however, this region is still inside the cluster tidal radius ($r_t \approx 16 \arcmin$).}
The images consisted of multiple dithered exposures in $F275W$, $F336W$, and $F438W$, for a total exposure time of 1,810s, 592s, and 217s, respectively. They were reduced by following the same technique used to produce the HUGS catalog. We refer the reader to \cite{Bellini2017} and \cite{Nardiello2018} for details on the data reduction method.


The photometric catalogs include 
145,610 and 10,897 sources with a measurement in the $F275W$ and $F336W$ filters, and brighter than $F275W$ = 27~mag, for the internal and external field, respectively. 
The $F275W,\ F275W - F336W$ color-magnitude diagrams (CMDs) including all stars from both catalogs are shown in the left panel of Fig.~\ref{336CMD Intro}. The photometry is very deep, reaching $F275W \approx$ 25 with a Signal-to-Noise ratio $S/N \approx$ 5. 
{ The extended blue horizontal branch (HB) is visible in both fields, and numerous blue straggler stars (BSS) are observed in the internal field. We counted them, and there are 342 and 8 BSSs in the internal and external field, respectively. Since the internal field has, on average, about 13 times more sources than the external one, these numbers indicate an excess of BSSs in the core of NGC~2808. This result is not surprising, since BSSs are thought to originate from binary star evolution and interaction, and their radial distribution is more peaked in the cluster cores \citep{2004ApJ...603..127F}.}

The HB clearly separates in four groups, as we will discuss in Section~\ref{Analysis of the Horizontal Branch Stars}. 
The cluster red-giant branch (RGB) splits in multiple sequence in this CMD, since the $F275W$ and $F336W$ filters are sensitive to the different light-element abundance of the stars \citep{johnson23}. 
A well-defined WD cooling sequence is visible and extends from $F275W \approx$ 21 down to 26 mag and has a $F275W - F336W$ color in the range $\approx$ -1.0 -- 0.5; a few objects in between the HB and the brightest WDs are present in the internal field and will be discussed in Section~\ref{section:analysis}.
In order to verify that the WD cooling sequence does not include cosmics or image artifacts, we used the drizzled stacked images provided by the HUGS survey to create a three-color image of the field of view. We then generated cut-out images around each selected WDs and checked them by eye. All of them resulted to be real sources, fainter and bluer than most neighboring stars. 
The cut-outs for a sample of WDs of the internal field are shown in Appendix~\ref{wd_cutouts}.

Out of the 145,610 sources, 122,228 have a measured membership probability P$_\mu$ in the internal field, i.e. $\approx$ 84\% of the total.
When selecting sources with P$_\mu$ $>$ 80\%, 117,580 objects are left, i.e. $\approx$ 96\%. The contamination by field stars of the internal field catalog is then $\lesssim$ 5\%; therefore, we used the full catalog with no PM cuts for our analysis to avoid introducing systematic effects in the star count calculations.
No WD in this catalog has a membership probability P$_\mu$ value; these faint blue stars do not have a PM measurement because most of them were not identified in the optical ACS images, which provide the first epoch for the PMs. 
Therefore, we downloaded the more recent catalogs of PMs for NGC~2808 from the HACKS { \citep{https://doi.org/10.17909/jpfd-2m08}} survey\footnote{\href{https://archive.stsci.edu/hlsp/hacks}{https://archive.stsci.edu/hlsp/hacks}}, based of multiple epochs of WFC3 and ACS data \citep{libralato2022}. By matching this catalog with the HUGS photometry we only found 21 WDs with PM measurements. As a consequence, we did not apply any PM cuts to  WDs and assumed that all the WD stars in the selected color and magnitude range of the $F275W$ vs. $F275W$ - $F336W$ CMD are cluster members. 
There might be a residual contamination by field stars on the WD cooling sequence but we expect it to be minor ($\lesssim$ 5\%) since the very high level of crowding of the cluster core and the majority of field stars being redder and closer compared to the WDs, with the consequence they do not overlap the CMD \citep{cal08mainpaper}. 

For the external field, we have no PM measurements and we did not apply any membership cut. All the sources from this catalog are plotted in the CMDs in the right panel of  Fig.~\ref{336CMD Intro}. Recently, \cite{Griggio2025} measured PMs for this field and 80\% of the observed WDs are also included. By selecting only sources with a PM in both the RA and DEC direction $\le\pm$ 5 mas/yr (the cluster PM is centered on 0) $\approx$ 96\% of the MS stars are retained, and all the WDs, further supporting our choice of retaining all measured stars in the internal field, where field star contamination should be comparable or less due to the higher crowding.

Summarizing, we did not apply any PM selection on both catalogs to avoid introducing any bias into our analysis.  

\subsection{Artificial Star Test and Completeness}\label{AST}

The Artificial Star Test (AST) for the MS and RGB phases were run separately from the test on the WD cooling sequence for both fields. We generated a synthetic CMD, excluding the WD cooling phase, in all the WFC3-UVIS and ACS filters, including $\approx$ 160,000 stars distributed approximately following a King profile. We added these stars one at a time to the individual images, considering the positional and photometric transformations reported in \cite{Nardiello2018} or adopted for the external field. We measured the single artificial star using the same procedure adopted for the real stars, and finally removed the artificial star from the images, to add the next one. 

We generated a synthetic catalog for the WD cooling sequence in all the WFC3-UVIS and ACS filters, by using a 0.54 $M_{\odot}$ CO-core cooling track, and including $\approx$ 64,000 sources, with the same spatial distribution as for the other AST. The test was run by following the same method. 

\begin{figure*}
    \centering
    \includegraphics[width=1\linewidth]{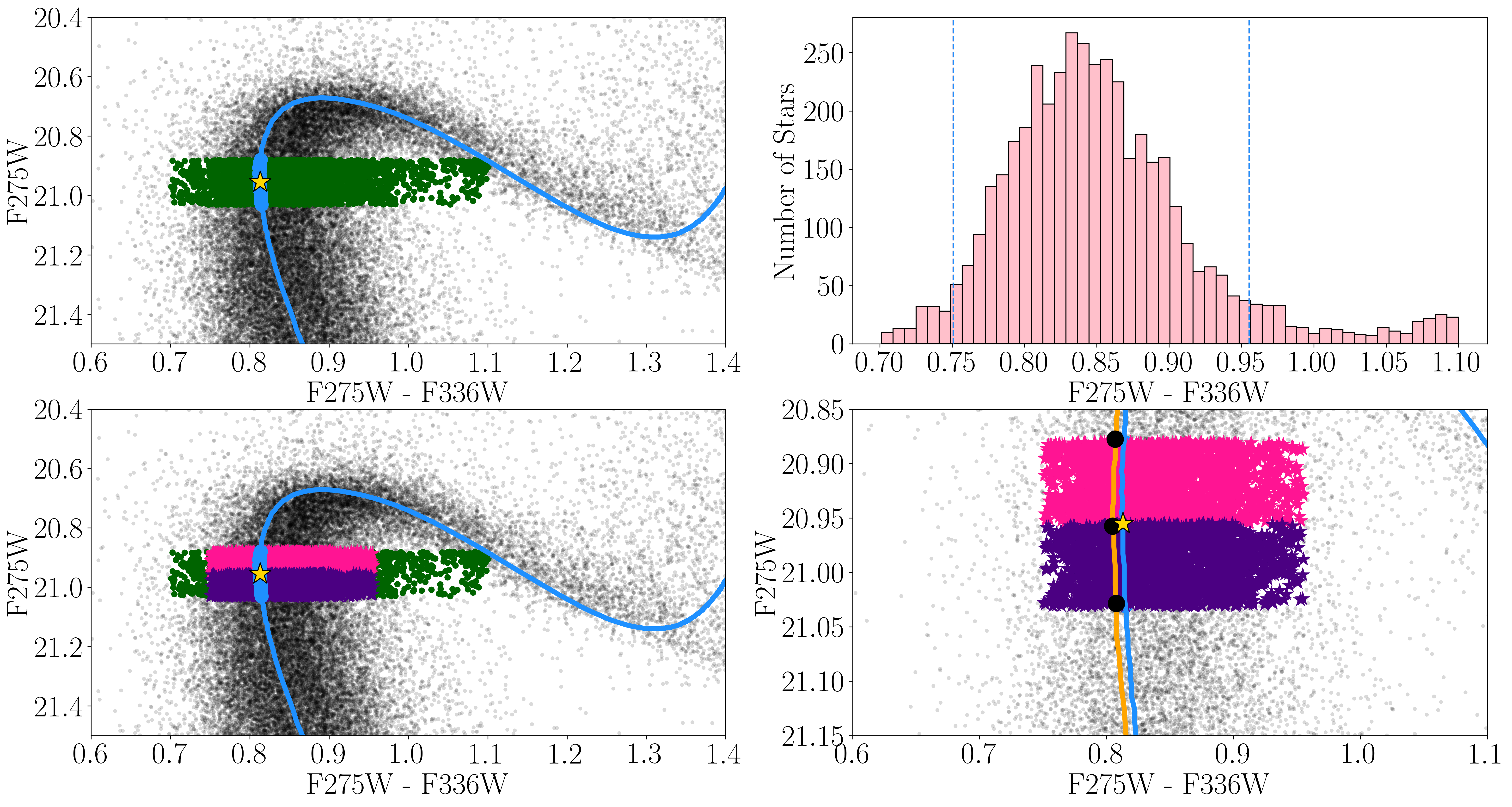}
    \caption{Top-left -- $F275W,\ F275W - F336W$ CMD of the internal field zoomed in around the MSTO region for all sources with an error in magnitude of $F275W$ and magnitude of $F336W$ $< 1$. A BaSTI $\alpha$-enhanced isochrone for $t$ = 12 Gyr, $Z$ = 0.002, and $Y$ = 0.248, is overplotted as a solid light blue line. The model was transformed to the observed plane by using the reddening and extinction values listed in Section~\ref{MSTO Analysis}. The theoretical MSTO at $M$ = 0.8 $\text{M}_\odot$ is marked with a golden star. 
    { The thicker section of the isochrone indicates a mass width $\delta\mathcal{M} = \pm 0.004$ around 0.8 M$_\odot$ of 20.88  $\le $F275W$ \le$ 21.03.}
    Stars selected in this MSTO region are marked as green dots. Top-right -- Color histogram of the MSTO selected stars. The dotted blue vertical lines mark the $\pm 1.5\sigma$ color cut used to select the final sample of MSTO stars. Bottom-left -- Final sample of MSTO stars in the same CMD. The $F275W$ magnitude bin is split in half, with the stars in the upper (N = 1,970) and bottom bin (N = 1,902) marked with pink and purple dots, respectively. Stars discarded by the color selection are marked with green dots. Bottom-right -- A BaSTI evolutionary track for $M$ = 0.8 $\text{M}_\odot$ (solid orange line) and the same isochrone (solid light blue) are overplotted on the MSTO region of the CMD. The { filled-black circles} on the track mark the 
    $F275W$ magnitudes used to calculate the MSTO crossing times (see Table~\ref{Table MSTO} for more details).}
    \label{MSTO Plot}
\end{figure*}

\begin{deluxetable*}{cc|ccc|cc}
\tablecaption{Star count-crossing time analysis for MSTO stars.}
\tablehead{
\colhead{Field} & \colhead{Bin} & \colhead{Count} & \colhead{Count*} & \colhead{Crossing Times (Myrs)} & \colhead{Total Star Count*} & \colhead{Total Crossing Times (Myrs)}
}
\startdata
\multirow{2}{*}{Internal} & Upper & 1970 $\pm$ 44 & 2082 $\pm$ 46 & 354 & \multirow{2}{*}{4098 $\pm$ 64} & \multirow{2}{*}{687 $\pm$ 69} \\
                          & Lower & 1902 $\pm$ 44 & 2016 $\pm$ 45 & 333 &                                &                             \\
\multirow{2}{*}{External} & Upper & 95 $\pm$ 10   & 105 $\pm$ 10  & 354 & \multirow{2}{*}{208 $\pm$ 14}  & \multirow{2}{*}{687 $\pm$ 69} \\
                          & Lower & 92 $\pm$ 10   & 102 $\pm$ 10  & 333 &                                &                             \\
\enddata
\tablecomments{Star counts and crossing times of MSTO stars for $\delta\mathcal{M} = 0.004$ values in the $F275W,\ F275W - F336W$ CMD have been presented. For both the fields, only all sources with an error in magnitude of $F275W$ and magnitude of $F336W$ $< 1$mag were selected. Note that asterisk-marked star counts are accounted for completeness. In the last two columns, we present the combined star counts and crossing times for the respective fields and bins. The errors in star counts were calculated considering them Poissonian.}
\label{Table MSTO}
\end{deluxetable*}

To derived the completeness as a function of the $F275W$ magnitude in the $F275W,\ F275W - F336W$ CMD, we analyzed the output catalogs of the AST in the following way: only stars with x and y $\le$ 1 pixel in distance from the input star, and with a $\delta$ mag $\le$ 0.75 mag in $F275W$ and $F336W$ were selected; the ratio of input over output stars provides the catalog completeness. The smoothed completeness function ($\phi$) is plotted versus $F275W$ in Fig.~\ref{AST Curves}, for both the internal and external field. A completeness level of 50\% is reached at $F275W \approx$ 26 and 28~mag for MS stars in the internal and external field, respectively, while the WD cooling sequence has 50\% completeness at $F275W \approx$ 24.3 and 25.7 mag. 

\section{Method and Analysis}\label{section:analysis}

In this investigation, we used a similar method as the one adopted by \citet{2007Castellani} and \citet{cal08mainpaper} to compare star counts and theoretical lifetimes of stars in different evolutionary phases. 
The number of stars in an advanced evolutionary stage, i.e. post MSTO, is proportional to the lifetime spent in that phase or region of the CMD; a deviation from this trend would suggest an excess or deficiency of stars observed.

The presence of SCWDs or He-core WDs in NGC~2808 would result in an excess of stars on the WD sequence, due to the slower cooling time of these WDs compared to canonical CO-core WDs.
Therefore, to investigate for the presence of SCWDs or He-core WDs, we compared star counts and evolutionary times along the WD cooling sequence of NGC~2808 to star counts and lifetimes of a region 
around the MS turn-off (MSTO).
The number of stars along the MS of a GGC is not proportional to the lifetime spent on this evolutionary phase, since there is a strong dependence on the initial mass function; however, to minimize this effect we selected a limited magnitude region ($\leq$ 0.15 mag) around the MSTO; in this case, stars have approximately the same mass and their number is minimally affected by the initial mass function \citep{2000Zocalli}. 

\subsection{Main-sequence turn-off stars}\label{MSTO Analysis}


In order to define the typical stellar mass of MSTO stars we used an $\alpha$-enhanced isochrone from the BaSTI database \citep{2021Pietrinferni}\footnote{http://basti-iac.oa-abruzzo.inaf.it/index.html} with $t$ = 12 Gyr and a metallicity of $Z$ = 0.002 for NGC~2808 (see Table~\ref{Table NGC~2808 Parameters}). 
We used the \cite{1989Cardelli} reddening law to calculate the extinction values $\text{A}_{\text{$F275W$}} = 1.799\times\text{A}_{\text{V}}$ and  $\text{A}_{\text{$F336W$}} = 1.643\times\text{A}_{\text{V}}$, with $E(B-V)$ = 0.19, and assumed a distance modulus of $\mu_{0}$= 15.05 mag (Table~\ref{Table NGC~2808 Parameters}) to overplot the isochrone on the CMD (solid blue line in Fig.~\ref{MSTO Plot}). The plot shows a very good agreement between theory and observations and enabled us to identify the TO point, i.e. the (hottest) bluest point of the MS, and estimate the typical stellar mass of the MSTO to be 0.8 $\text{M}_{\odot}$.


{ We used an MSTO mass width of $\pm$ 0.004 M$_\odot$ to define the brightness width of the MSTO.}
The $F275W$ magnitude of the points along the isochrone of mass 0.8 $\pm$ 0.004 $\text{M}_\odot$ correspond to a magnitude difference of 0.13. The selected stars in the magnitude 20.88 $< F275W <$ 21.03 and color 0.7 $< F275W - F335W <$ 1.1 range are shown in the $F275W$, $F275W$ - $F336W$ CMD in the top left panel of Fig.~\ref{MSTO Plot} for the internal field.


{ We selected stars within 1.5 sigma of the mode of the main sequence color distribution to maximize star counts while ensuring rejection of sub-giant branch (SGB) stars.}


The MSTO region is further divided in two magnitude bins referred to as the {upper bin} (UB, pink dots in the bottom panels of Fig.~\ref{MSTO Plot}) and the {lower bin} (LB, purple). We applied this magnitude split to investigate the dependence of the results on the adopted distance modulus for NGC~2808 (see Table~\ref{Table NGC~2808 Parameters} and Table~\ref{tab:MSTO numbers with various delta alpha values}).
The star counts in the MSTO region are corrected for completeness by using the AST results from Section ~\ref{AST}.

To calculate the evolutionary lifetime across the MSTO magnitude bin for a star of mass 0.8 $\text{M}_\odot$, an $\alpha$-enhanced evolutionary evolutionary track for this mass and appropriate chemical composition for NGC~2808 is used. The crossing time is calculated as the time difference between the magnitude limits of the MSTO upper and lower bins on the track. Table~\ref{Table MSTO} lists the MSTO crossing times for the two bins.

{ The same analysis was performed for the external field CMD.}
To test the effect of differential reddening on the MSTO star counts, we re-calculated them after correcting the CMD of the internal field for differential reddening (see Appendix~\ref{reddening} for more details).
{ However, the differential extinction across the field was found to be of the same order of magnitude as the interpolation uncertainty between grid points and had no significant impact on the results.}
Additionally, various $\delta\mathcal{M}$ values (see Appendix~\ref{section:deltaalphavalues}) and isochrones of different ages were used to test the dependence from these input parameters. Table~\ref{Table MSTO} only lists star counts obtained by using $\delta\mathcal{M} = $ 0.004 M$_\odot$ and for an age of 12 Gyr.

\subsection{White Dwarf stars}\label{WD Analysis}

\begin{figure*}
    \centering
    \includegraphics[width=0.7\linewidth]{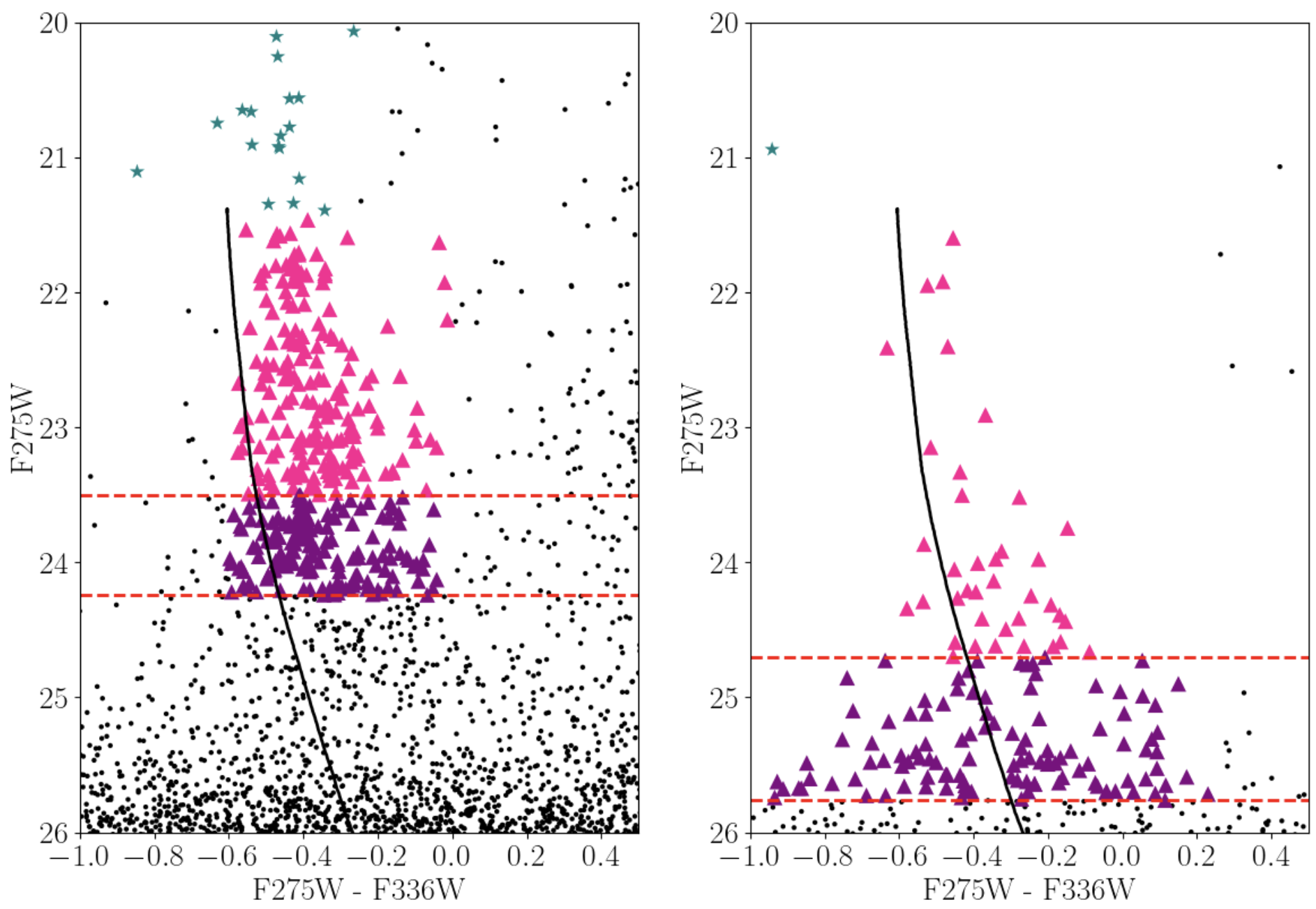}
    \caption{Left -- $F275W,\ F275W - F336W$ CMD of the internal field zoomed around the WD cooling sequence region. Stars marked as teal asterisk-shaped stars are the objects excluded from this study. Pink and purple triangles mark selected WDs divided into magnitude bins (see text for more details). The dashed horizontal lines indicate the lower bounds of the magnitude bins, and the solid black line is a BaSTI CO-core/H-envelope WD cooling model for a mass of 0.54 $\text{M}_\odot$. Right -- Same zoomed-in CMD for the external field. Light and dark blue triangles mark selected WDs in the two magnitude bins.} 
    \label{WD Plot}
\end{figure*}


\begin{deluxetable*}{cccc|cc}
\tablecaption{Star count-cooling time analysis for WD stars.}
\tablehead{
\colhead{Field} & \colhead{Bin} & \colhead{$\text{m}_{F275W}$} & \colhead{Counts} & \colhead{Counts*} & \colhead{Cooling Times (Myrs)}
}
\startdata
Internal & Bright & 23.5 & 181 $\pm$ 13 & 228 $\pm$ 15 & 13 $\pm$ 1 \\
Internal & Faint  & 24.3 & 336 $\pm$ 18 & 489 $\pm$ 22 & 28 $\pm$ 3 \\
External & Bright & 24.7 & 38 $\pm$ 6   & 43 $\pm$ 7   & 52 $\pm$ 5 \\
External & Faint  & 25.8 & 152 $\pm$ 12 & 223 $\pm$ 15 & 165 $\pm$ 16 \\
\enddata
\tablecomments{WD star counts and cooling times from the $F275W,\ F275W - F336W$ CMD are listed for the internal and external fields. Asterisk-marked counts are completeness corrected. The listed $\text{m}_{\text{$F275W$}}$ are the faintest magnitudes of the bins in which the WDs were selected. The `bright bin' corresponds to the brighter part of the WD cooling sequence, while the `fainter bin' corresponds to the entire WD sequence until the magnitude for which completeness is 50\%.}
\label{Table WD}
\end{deluxetable*}


Fig.~\ref{WD Plot} shows the CMD zoomed-in the WD region of the internal (left panel) and external (right) field. The typical stellar mass at the MSTO is 0.8 $\text{M}_\odot$, and the theoretical initial-to-ﬁnal mass relationship predicts masses of $\approx$ 0.53–0.55 $\text{M}_\odot$ \citep{WeissFerguson2009} for newly formed WDs. This prediction is observationally supported by the spectroscopic measurements of bright WDs in the GGC M~4 by \cite{kalirai07}, for which they find a mean mass of $\approx$ 0.53 $\text{M}_\odot$. We used a BaSTI cooling track for DA (hydrogen envelope) CO-core WDs with $M =$ 0.54 $\text{M}_\odot$ to fit the WD cooling sequence of NGC~2808 \citep{2022Salaris}, by using the reddening and distance modulus of Table~\ref{Table NGC~2808 Parameters} and the same extinction coefficients used to fit isochrones to the MS and RGB of the cluster.


In the case of the internal field, which covers the core of NGC~2808, the cooling track is systematically bluer compared to the WD sequence (see left panel of Fig.~\ref{WD Plot}).
On the other hand, the $\alpha$-enhanced isochrone for $t$ = 12 Gyr, $Z$ = 0.002, and He-canonical, $Y$ = 0.248, is in very good agreement with the observations, within uncertainties, and shows no systematic shift in color (see Fig.~\ref{MSTO Plot}).
To find the culprit, we analyzed the AST results by comparing the input and output color of the recovered stars as a function of the $F275W$ magnitude. A systematic color shift increasing at fainter magnitudes was found, with values of $\approx$ 0.2~mag at the bottom of the detected WD sequence. 

{ This result points towards crowding affecting the measured magnitude of the WDs. Since most of the stars in a cluster belong to the MS, they are systematically redder than WDs; if the measurement of a WD is affected by a neighboring MS star, its $F275W$ magnitude will be fainter, and so its $F275W - F336W$ color would become redder. 
The effect is more pronounced in the inner field photometric catalog, showing a maximum color shift of 0.2 mag, compared with the external field, where the maximum shift is 0.03 mag relative to the cooling sequence (Fig.~\ref{WD Plot}).
Therefore, the color shift between the model and the cooling sequence does not affect our star count analysis.}


\begin{figure*}
    \centering
    \includegraphics[width=0.7\linewidth]{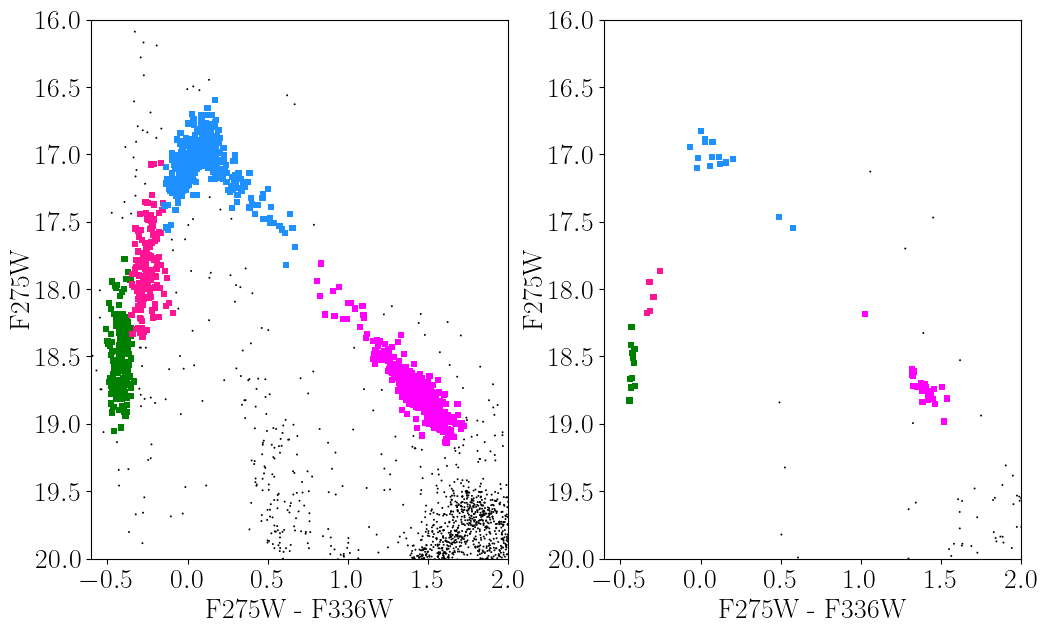}
    \caption{Left -- Zoom-in around the HB region of NGC~2808 $F275W$ vs. $F275W - F336W$ CMD of the internal field. The HB is divided into four groups: the red horizontal branch (RHB, purple) and three extended blue tails, namely, EBT1 (cyan), EBT2 (magenta), and EBT3 (green). Right -- Same but for the external field. { We refer the reader to \cite{Castellani2007HB2808}} for method of division of HB into four parts.}
    \label{HB CMD Counts}
\end{figure*}

\begin{deluxetable*}{ccclc}
\tablecaption{Star count-evolutionary time analysis for HB stars.}
\tablehead{
\colhead{HB Populations} & 
\colhead{Internal Field} & 
\colhead{External Field} & 
\colhead{$\langle \text{M}_{\text{HB}} \rangle$ (M$_\odot$)} & 
\colhead{Evolutionary Times (in Myrs) [canonical]}
}
\startdata
RHB   & 577 $\pm$ 24 (43\% $\pm$ 3\%)  & 27 $\pm$ 5 (44\% $\pm$ 14\%)  & 0.690 & 101 $\pm$ 10 \\
EBT1  & 474 $\pm$ 22 (35\% $\pm$ 3\%)  & 16 $\pm$ 4 (26\% $\pm$ 10\%)  & 0.565 & 112 $\pm$ 11 \\
EBT2  & 139 $\pm$ 12 (10\% $\pm$ 1\%)  & 5 $\pm$ 2 (8\% $\pm$ 5\%)     & 0.479 & 129 $\pm$ 13 \\
EBT3  & 154 $\pm$ 12 (11\% $\pm$ 1\%)  & 13 $\pm$ 4 (21\% $\pm$ 9\%)   & 0.490 & 127 $\pm$ 13 \\
\hline
Total & 1344 $\pm$ 37                 & 61 $\pm$ 8                   & $\cdots$ & $\cdots$ \\
\enddata
\tablecomments{HB star counts and relative fractions with their uncertainties from the $F275W,\ F275W - F336W$ CMD for the internal and external fields.
Evolutionary times for the four HB groups and average HB masses, ${\text{M}_{\text{HB}}}$, are also listed. The errors in star counts were calculated considering them Poissonian.}
\label{Table HB}
\end{deluxetable*}

The CMD of the internal field shows about a dozen stars (marked as teal star symbols in Fig.~\ref{WD Plot}) which are brighter than $F275W \approx$ 21.4~mag, i.e., the brightest point of the cooling track. These stars could be EHBs in their descent phase towards the WD cooling sequence (since they skip the AGB phase) or they could be cataclysmic variables (CVs) or dwarf novae (DNe). 
{ There is just one such star in the external field.}
These stars were not included in our analysis, and we defer a more detailed investigation to a future paper.


No photometric or PM cuts were applied to select WDs in both fields. We set the upper limit of the WD bin selection at $F275W \approx$ 21.4~mag, i.e. the brightest point of the cooling track, and the fainter limit is set by the $\approx$ 50\% completeness level at $F275W \approx$ 24.3 mag for the internal field, and $\approx$ 25.8~mag for the external one (see the dashed horizontal lines in Fig.~\ref{AST Curves}). For the internal field WDs, we produced ``WD cutouts'' which are postage stamp images of the WDs. In this way, we were able to confirm the validity of the WD. 
We refer the reader to Appendix~\ref{wd_cutouts} for more details. 
Further, the WD sequence of each field was divided into two magnitude bins, with 23.5 and 24.7~mag marking the magnitude that separates the two for the internal and external field, respectively. All the stars above this magnitudes belong to the `bright bin', while those up to 24.3 and 25.8~mag, in the internal and external fields, respectively, belong to the `faint bin'. This division is done in order to verify if results are independent from the selected faint magnitude limit and completeness (even though we correct for it). 
The WD star counts for each bin are calculated and corrected for completeness and listed in Table~\ref{Table WD}. The faint magnitude limit of the WD bins was then used to determine their cooling times, listed in the same table. 

\subsection{Horizontal Branch stars}\label{Analysis of the Horizontal Branch Stars}


NGC~2808 has an extended and well-populated HB, composed of two main parts, the red horizontal branch (RHB) and the extended blue tail (EBT). The EBT can be further divided in three parts, namely, EBT1, EBT2, and EBT3, where EBT3 is the bluest on the $F275W,\ F275W - F336W$ CMD, following \citet{Castellani2007HB2808}. Fig.~\ref{HB CMD Counts} shows a zoom-in around the HB region of the CMD for the internal field (left panel) and the external one (right). Stars belonging to the four groups are highlighted with different colors in the plot. 

We selected HB stars from the $F275W,\ F275W - F336W$ CMD without applying a proper motion selection since most field stars in this CMD are redder than the HB, resulting in a minimal contamination ($<$ 10\%, \citealt{2007Castellani}). 
We did not run an AST along the HB and we were not able to apply a completeness correction to the star counts; however, HB stars are systematically brighter than the RGBs in $F275W$, with magnitudes in the range 16.5 $\lesssim F275W \lesssim$ 19~mag, where the photometric completeness is above 90\% in both fields (see Figs.~\ref{AST Curves} and \ref{HB CMD Counts}). 
Table~\ref{Table HB} lists the HB star counts for the four groups. 

To derive the HB lifetimes we used the evolutionary tracks for the mean mass of stars in the four groups ($\langle{\text{M}_{\text{HB}}}\rangle$), NGC~2808 metallicity, $Z = 0.002$, and a He-canonical abundance, Y = 0.248. The average mass values for the four HB groups were taken from Table 1 of \cite{Dalessandro2011}, and are listed in Table~\ref{Table HB}, together with the lifetimes.




To find the average lifetime for the entire HB, we first derived the fraction of stars in each HB group, as listed in parentheses in Table~\ref{Table HB}. Further, we normalized the lifetimes of each HB group by the corresponding population fraction. For instance, for the internal field, the relation is given in Equation~\ref{HB Total Time}, where $T$ is the average HB lifetime, and $t_{\text{RHB}}, t_{\text{EBT1}}, t_{\text{EBT2}}, t_{\text{EBT3}}$ are the lifetimes for the different groups. 
The final HB evolutionary lifetimes is 111 $\pm$ 11 Myrs:

\begin{equation*}
    T = [0.43\times t_{\text{RHB}}] + [0.35\times t_{\text{EBT1}}]
\end{equation*}

\begin{equation}\label{HB Total Time}
    \hspace{0.05cm} + [0.10\times t_{\text{EBT2}}] + [0.11\times t_{\text{EBT3}}]
\end{equation}

\begin{figure*}
    \centering
    \includegraphics[width=0.7\linewidth]{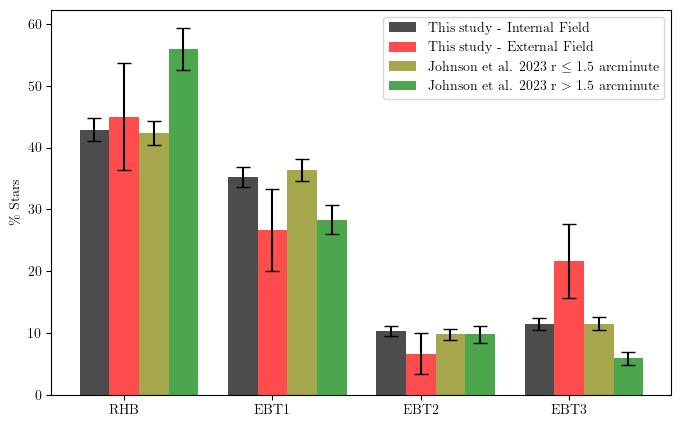}
    \caption{Histogram of star counts of internal field and external and from \cite{johnson23}. Error bars have been plotted assuming a Poissonian distribution.}
    \label{HB Histogram}
\end{figure*}

To validate our HB population fractions, we compared them to the ones found by \cite{johnson23};
their work reports HB star counts for the same four HB groups for the core region, i.e. for distances from the center $r \le$ 1.5\arcmin, and for the rest of the cluster. Their analysis is based on wide-field DECam@4m-Blanco photometry covering the entire extension of NGC~2808, and the same HST HUGS data for the cluster core. The relative fractions of stars in the four HB groups with errors from \cite{johnson23} are listed in Table~\ref{Table Johnson and Our Study HB Counts}, while the fractions from both studies are shown in Fig.~\ref{HB Histogram}; this plot shows that the agreement between the two analysis is very good, within uncertainties. 


In both analyses, the RHB star fraction remains constant when moving from the center to the cluster outer regions (with a small overabundance in \cite{johnson23} possibly due to field star contamination); EBT2 star distribution is also flat across NGC~2808, while the EBT1 stars decrease at larger distances from the cluster core in both studies. This could be due to them originating from a $He$-enhanced population in NGC~2808 or the 'hot-flasher' scenario (see Section 5.1 of \cite{johnson23} for a detailed discussion). It is interesting to note how the EBT3 stars decrease in \cite{johnson23} and increase in our analysis, with increasing distance from NGC~2808 core; even assuming field objects contaminate our EBT3 group, the contamination cannot be larger than $\approx$ 10\%, and the excess of these HB stars would persist. A possible culprit for the difference in the two studies could be that the \cite{johnson23} EBT3 group is affected by incompleteness at these faint magnitudes (20 $\lesssim r \lesssim$ 22 mag). No AST and completeness correction was applied to the DECam photometry.

\begin{deluxetable*}{ccc}
\tablecaption{HB star counts from \citet{johnson23}.}
\tablehead{
\colhead{HB Populations} & 
\colhead{r $\leq$ 1.5$\arcmin$} & 
\colhead{r $>$ 1.5$\arcmin$}
}
\startdata
RHB   & 484 $\pm$ 22 (42\% $\pm$ 3\%) & 274 $\pm$ 17 (56\% $\pm$ 6\%) \\
EBT1  & 416 $\pm$ 20 (36\% $\pm$ 3\%) & 139 $\pm$ 12 (28\% $\pm$ 4\%) \\
EBT2  & 112 $\pm$ 11 (10\% $\pm$ 1\%) & 48 $\pm$ 7 (10\% $\pm$ 2\%)   \\
EBT3  & 132 $\pm$ 11 (12\% $\pm$ 1\%) & 29 $\pm$ 5 (6\% $\pm$ 1\%)    \\
\hline
Total & 1144 $\pm$ 34               & 490 $\pm$ 22                  \\
\enddata
\tablecomments{Star counts and corresponding relative fractions of HB populations from \citet{johnson23}. The errors in star counts were calculated considering them Poissonian. See Fig.~\ref{HB Histogram} for the corresponding histogram.}
\label{Table Johnson and Our Study HB Counts}
\end{deluxetable*}


\section{Helium-enhancement}\label{He-enhanced Crossing Times for HB and MSTO stars}

As discussed earlier, NGC~2808 might host stellar sub-populations with different Helium abundances; for instance, the MS is split in three sequences, with the bluest one likely the most $He$-enhanced \citep{Piotto2007, Milone2012, Milone2015quintuple}. The helium enrichment could also explain the observed HB morphology \citep{dantona04, dantona05, Lee05}. Therefore, in this section, we derived crossing times for the MSTO phase, and evolutionary times along the HB by using $He$-enhanced stellar models.

\subsubsection{MSTO Stars} \label{He-enhanced-MSTO}

\begin{figure}[ht!]
    \centering
    \includegraphics[width=1\linewidth]{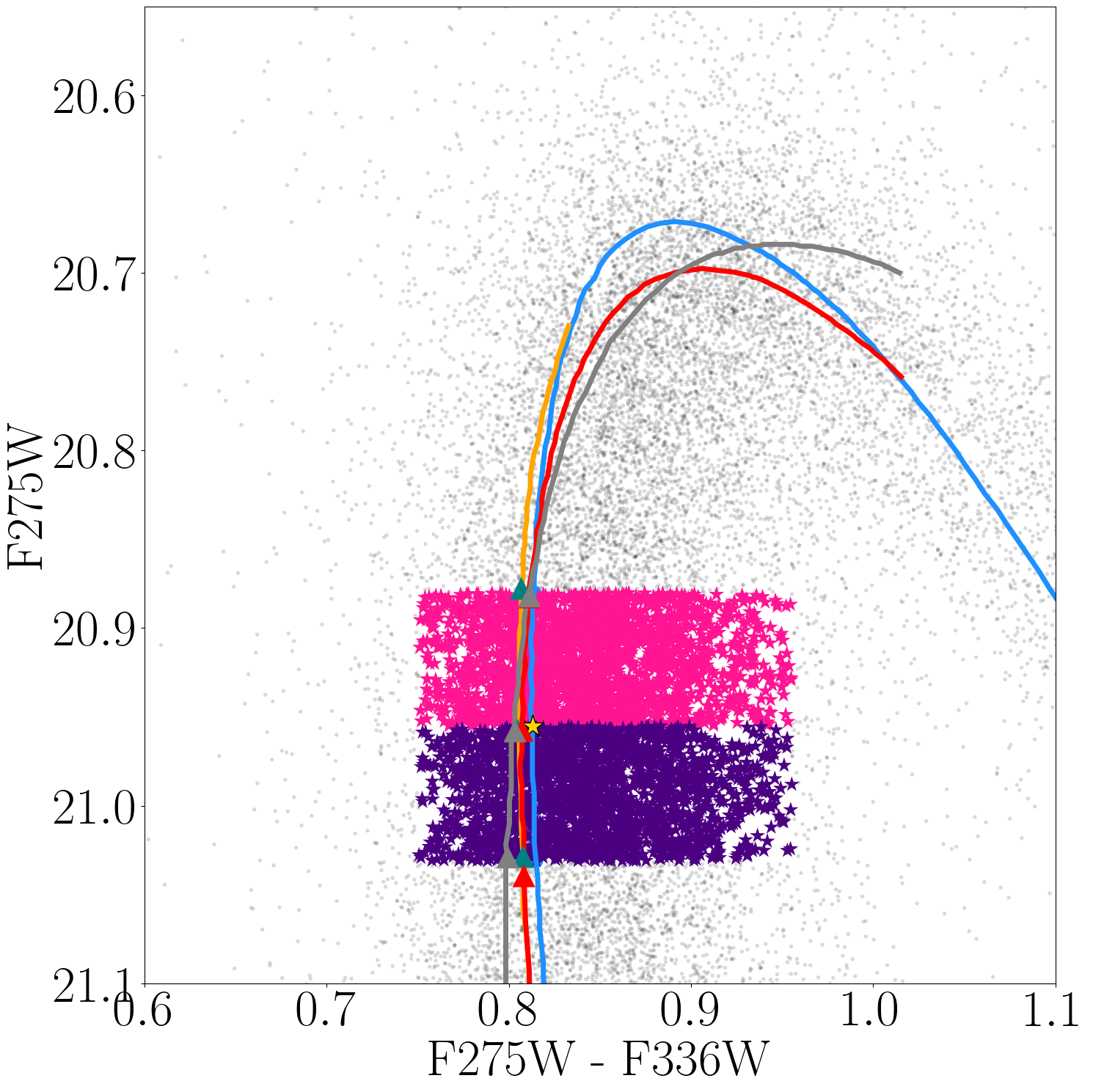}
    \caption{Plot is the same as the bottom-right CMD of Fig.~\ref{MSTO Plot}. The evolutionary track of mass 0.6 M$_\odot$ (in grey) and 0.7 M$_\odot$ (in red) has been plotted with Y = 0.400 and Y = 0.300, respectively. The grey and red triangles mark the point on the evolutionary track closest to the limit of the magnitude bin.}
    \label{MSTO Int Enhanced CMD}
\end{figure}


We derived $He$-enhanced crossing times for the MSTO region by assuming, for the $He$-enhanced sub-populations, the same age, $t$ = 12 Gyr, and metallicity, $Z = 0.002$, of the primordial stellar population, and Helium abundances of $Y$ = 0.30 and 0.40.  The typical masses at the MSTO are $M$ = 0.7 and 0.6~M$_\odot$, and crossing times are 629 $\pm$ 63 and 416 $\pm$ 42~Myr for $Y$ = 0.30 and 0.40, respectively.

We used the numbers from \cite{Milone2015quintuple} to determine the fraction of stars belonging to the $He$-canonical (primordial) and the two $He$-enhanced sub-populations along the MS of NGC~2808.
To derive a crossing time for the MSTO region that takes into account a combination of $He$-canonical and enhanced stellar sub-populations, we used the following equation:



\begin{equation}\label{MSTO He-enhanced equation}
    T = 0.50\times t_{\text{Y=0.248}} + 0.31\times t_{\text{Y=0.300}} + {0.19}\times t_{\text{Y=0.400}}
\end{equation}

which provides a final crossing time for of 617$\pm$62 Myr.

\subsubsection{HB Stars} \label{He-enhanced-HB}
To take into account the effect of Helium enhancement along the HB of NGC~2808, we estimated evolutionary lifetimes by assuming that RHB is the progeny of the primordial stellar population of the cluster, and has a canonical Helium abundance. EBT1 are assumed to be $He$-enhanced, with  $Y$ = 0.300, following \citet{Dalessandro2011}. However, \citet{johnson23} recently showed that the RHBs might be the progeny of more than one stellar sub-population in the cluster, while EBT1 seem to be the progeny of the enriched sub-populations only (see also \cite{Gratton2011}). Stars in the EBT2 and EBT3 groups are usually hotter than $\approx$ 11,500 K (the so-called Grundahl u-jump) and have atmospheric abundances affected by radiative levitation and diffusion, so it is not possible to observe their original chemical abundances. 
{ For this reason, we assume EBT2 are He-enhanced, while EBT3 mostly originate from the hot-flashers remnants \citep{Moehler2004, Castellani2007HB2808, 2012Brown}.}
The total HB evolutionary time is estimated using Equation~\ref{HB Total Time}, and results to be 113$\pm$11 Myr.

\section{Comparing star counts to evolutionary lifetimes}\label{SC-CT}


\begin{deluxetable*}{ccc|ccccc}
\tablecaption{WD/MSTO Ratios.}
\tablehead{
\colhead{Field} & \colhead{Bin} & $\text{m}_{\text{$F275W$}}$ Limit & \colhead{$N_{\text{WD}}/N_{\text{MSTO}}$} & \multicolumn{2}{c}{He-canonical} & \multicolumn{2}{c}{$He$-enhanced} \\
\colhead{} & \colhead{} & \colhead{} & \colhead{} & \colhead{$t_{\text{WD}}/t_{\text{MSTO}}$} & \colhead{Excess WD \%} & \colhead{$t_{\text{WD}}/t_{\text{MSTO}}$} & \colhead{Excess WD \%}
}
\startdata
\multirow{2}{*}{Internal} & Bright & 23.5 & 0.056 $\pm$ 0.004 & 0.019 $\pm$ 0.003 & 66.4 $\pm$ 5.3 & 0.021 $\pm$ 0.003 & 62.6 $\pm$ 5.9 \\
 & Faint & 24.3 & 0.119 $\pm$ 0.006 & 0.041 $\pm$ 0.006 & 65.6 $\pm$ 5.1 & 0.046 $\pm$ 0.006 & 61.7 $\pm$ 5.7 \\
\multirow{2}{*}{External} & Bright & 24.7 & 0.209 $\pm$ 0.035 & 0.075 $\pm$ 0.011 & 64.0 $\pm$ 7.9 & 0.084 $\pm$ 0.012 & 59.9 $\pm$ 8.8 \\
 & Faint & 25.8 & 1.071 $\pm$ 0.103 & 0.241 $\pm$ 0.034 & 77.5 $\pm$ 3.8 & 0.268 $\pm$ 0.038 & 75.0 $\pm$ 4.3 \\
\enddata
\tablecomments{$N_{\text{WD}}/N_{\text{MSTO}}$ are the star count ratios and $t_{\text{WD}}/t_{\text{MSTO}}$ are the evolutionary time ratios for the internal and external fields. The ``$\text{m}_{\text{$F275W$}}$ limit'' is the faintest limit up to which the WDs have been selected in a particular bin.}
\label{Ratio Table WD MSTO}
\end{deluxetable*}


\begin{deluxetable*}{ccc|ccccc}
\tablecaption{WD/HB Ratios.}
\tablehead{
\colhead{Field} & \colhead{Bin} & $\text{m}_{\text{$F275W$}}$ Limit & \colhead{$N_{\text{WD}}/N_{\text{HB}}$} & \multicolumn{2}{c}{He-canonical} & \multicolumn{2}{c}{$He$-enhanced} \\
\colhead{} & \colhead{} & \colhead{} & \colhead{} & \colhead{$t_{\text{WD}}/t_{\text{HB}}$} & \colhead{Excess WD \%} & \colhead{$t_{\text{WD}}/t_{\text{HB}}$} & \colhead{Excess WD \%}
}
\startdata
\multirow{2}{*}{Internal} & Bright & 23.5 & 0.169 $\pm$ 0.012 & 0.115 $\pm$ 0.016 & 32.0 $\pm$ 10.7 & 0.114 $\pm$ 0.016 & 32.9 $\pm$ 10.6 \\
 & Faint & 24.3 & 0.364 $\pm$ 0.019 & 0.253 $\pm$ 0.036 & 30.4 $\pm$ 10.5 & 0.250 $\pm$ 0.035 & 31.3 $\pm$ 10.3 \\
\multirow{2}{*}{External} & Bright & 24.7 & 0.710 $\pm$ 0.141 & 0.463 $\pm$ 0.066 & 34.7 $\pm$ 15.9 & 0.458 $\pm$ 0.065 & 35.5 $\pm$ 15.7 \\
 & Faint & 25.8 & 3.648 $\pm$ 0.527 & 1.485 $\pm$ 0.210 & 59.3 $\pm$ 8.2 & 1.468 $\pm$ 0.208 & 59.8 $\pm$ 8.1 \\
\enddata
\tablecomments{$N_{\text{WD}}/N_{\text{HB}}$ are the star count ratios and $t_{\text{WD}}/t_{\text{HB}}$ are the evolutionary time ratios for the internal and external fields. The ``$\text{m}_{\text{$F275W$}}$ limit'' is the faintest limit uptil which the WDs have been selected in a particular bin.}
\label{Ratio Table WD HB}
\end{deluxetable*}

\begin{deluxetable*}{cccccc}
\tabletypesize{\scriptsize}
\tablecaption{HB/MSTO Ratios.}
\tablehead{
\colhead{} & \colhead{} & \multicolumn{4}{c}{Evolutionary Time Ratio} \\
\colhead{} & \colhead{} & \colhead{HB (EBT1\&2 canonical),} & \colhead{HB (EBT1\&2 enhanced),} & \colhead{HB (EBT1\&2 enhanced),} & \colhead{HB (EBT1\&2 canonical),} \\
\colhead{} & \colhead{$t_{\text{HB}}/t_{\text{MSTO}} = $} & \colhead{MSTO canonical} & \colhead{MSTO canonical} & \colhead{MSTO enhanced} & \colhead{MSTO enhanced}
}
\startdata
\multicolumn{2}{c}{~} & 0.16 $\pm$ 0.02 & 0.16 $\pm$ 0.02 & 0.18 $\pm$ 0.03 & 0.18 $\pm$ 0.03 \\
\hline
Field & Star Count & \multicolumn{4}{c}{Excess HB \%} \\
Internal & 0.33 $\pm$ 0.01 & 50.6 $\pm$ 7.1 & 50.0 $\pm$ 7.2 & 44.3 $\pm$ 8.0 & 45.0 $\pm$ 7.9 \\
External & 0.29 $\pm$ 0.04 & 44.8 $\pm$ 11.2 & 44.2 $\pm$ 11.3 & 37.9 $\pm$ 12.6 & 38.6 $\pm$ 12.4 \\
\enddata
\tablecomments{Star count–evolutionary time ratio for total HB and MSTO stars ($N_{\text{HB}}/N_{\text{MSTO}}$ and $t_{\text{HB}}/t_{\text{MSTO}}$).}
\label{Ratio Table MSTO HB 2}
\end{deluxetable*}




In order to investigate if an excess of bright WDs is present in NGC~2808, we compared the observed ratio of star counts on the selected region on the WD cooling sequence and the MSTO (see Section~\ref{section:analysis}), to the ratio of lifetimes for these evolutionary phases. If an excess of WDs (or a lack) of MS stars is present, the star count ratio would be larger than the lifetime ratio.
The excess in percentage is given by:

\begin{equation}\label{eqn WD percentage excess}
    \text{Excess \%} = \Bigg[ 1 - \frac{\text{R}_{\text{LT}}}{\text{R}_{\text{SC}}} \Bigg] \times 100
\end{equation}

where $R_{SC}$ is the ratio of the star counts and $R_{LT}$ is the ratio of the evolutionary lifetimes.

We assumed that errors in the star counts are Poissonian, and so $\differentiald N = \sqrt{N}$; the error on the star count ratio ${R}_{SC} = N_1/N_2$ will then be:  




\begin{equation}\label{Error In R_SC}
    {\differentiald \text{R}_{\text{SC}} = \text{R}_{\text{SC}}\sqrt{\frac{N_{\text{2}} + N_{\text{1}}}{N_{\text{2}}\cdot N_{\text{1}}}}}
\end{equation}


We assumed an error of 10\% for the evolutionary lifetimes \citep{2021Pietrinferni}; the error on the lifetime ratio
$R_{LT} = t_1/t_2$ is then:



\begin{equation}\label{Error In R_LT}
     {\differentiald \text{R}_{\text{LT}} = 0.14\text{R}_{\text{LT}}}
\end{equation}


Therefore, we can derive the error in the excess of stars in percentage as:

\begin{equation}\label{error in percentage excess}
    {\differentiald \text{(EXS)}\% = \frac{\text{R}_{\text{LT}}}{\text{R}_{\text{SC}}}\sqrt{0.14^2 + \Bigg[\frac{\differentiald\text{R}_{\text{SC}}}{\text{R}_{\text{SC}}}\Bigg]^2} \times 100}
\end{equation}




\subsection{Comparing WDs to MSTO Stars}\label{WD vs. MSTO Stars}

We present the star count ratios, $N_{\text{WD}}/N_{\text{MSTO}}$, and evolutionary time ratios as the ratio of the WD cooling time to the MSTO crossing time, i.e, $t_{\text{WD}}/t_{\text{MSTO}}$ in Table~\ref{Ratio Table WD MSTO}. We used values from Tables~\ref{Table MSTO} and \ref{Table WD} to calculate the star count and canonical evolutionary time ratios. Table~\ref{Ratio Table WD MSTO} shows that star count ratios are three times larger than the evolutionary time ratios, for both fields. This discrepancy decreases to a factor of 2.5 when considering $He$-enhanced evolutionary times for a fraction of the MSTO stars. Note that the star count ratio for the fainter WD cooling sequence bin of the external field is systematically larger compared to the others; this could be due to a larger contamination of spurious objects towards the faint magnitude limit considered ($F275W$ = 25.8 mag).


Therefore, our analysis suggests an excess of WDs in NGC~2808, compared to what predicted by evolutionary models, ranging between $\approx$ 65 and 60\% on average, depending if considering only a $He$-canonical stellar population or a mix of canonical and $He$-enhanced stellar sub-populations in the cluster.
This excess is observed in both fields (internal and external), implying no dependence on the distance from the cluster center up to $\approx$ 5$'$, i.e. 5 times the cluster half-mass radius.


\subsection{Comparing WDs to HB Stars}\label{section HB vs. WD Stars}


We derived the ratio of star counts observed in the same region of the WD cooling sequence and stars on the HB for both fields, and compared these to the ratio of the WD cooling times and the HB evolutionary lifetimes. This was calculated for a canonical stellar population and a mix of canonical and $He$-enhanced sub-populations in Section~\ref{Analysis of the Horizontal Branch Stars} and Section~\ref{He-enhanced-HB}, respectively. The ratios are listed in Table~\ref{Ratio Table WD HB}.

Even in this case, the observed ratios are larger than the evolutionary time ratios, leading to an excess of WDs or to a lack of HB stars. Note that in this case also the observed ratio for the faint bin of the WD cooling sequence of the external field is larger compared to the other values, probably due to contamination.
The observed ratios are $\approx$ 30\% larger compared to the theoretical ones, and there is no difference between the central field and the external one, pointing to no radial dependence. Also, theoretical ratios are quite similar when accounting for a mix of canonical and $He$-enhanced stellar sub-populations; so assuming for the presence of $He$-enhanced sub-populations does not affect these findings.

\subsection{Comparing HBs to MSTO Stars}\label{section HB vs. MSTO Stars}

In order to verify if the observed ratio $N_{WD}/N_{HB}$ is larger than the predicted one for an excess of WDs or a lack of HB stars, we compared the total HB star counts to the total MSTO region star counts. Since we used canonical stellar population models and a mix of canonical and $He$-enhanced stellar population models for both MSTO and HB, we estimated four different combinations of evolutionary time ratios, $t_{HB}/t_{MSTO}$, all listed in Table~\ref{Ratio Table MSTO HB 2}. The corresponding $N_{\text{HB}}/N_{\text{MSTO}}$ are consistently larger than $t_{\text{HB}}/t_{\text{MSTO}}$, suggesting an excess of HB stars. 
{ The HB excess is $\approx$ 5\% larger in the internal compared to the external field, which might suggest a radial dependence of the ratio. However, the observed ratios of the two fields are still compatible within uncertainties (see Table~\ref{Ratio Table MSTO HB 2}).}

\section{Discussion}\label{section:discussion}


\begin{deluxetable*}{cccccc}
\tabletypesize{\normalsize}
\tablecaption{Various Parameters for Clusters which have been tested to host SCWDs.}
\tablehead{
\colhead{Cluster} & \colhead{Host SCWDs} & \colhead{Mass ($\times 10^{5}$ M$_{\odot}$)} & \colhead{[Fe/H]} & \colhead{$\log{(\rho_{\text{hmr}})}$ (M$_{\odot}$/pc$^3$)} & \colhead{$\log{(\text{t}_{\text{r}_{\text{c}}})}$ (years)}
}
\startdata
M3 & No & 4.09 & -1.5 & 2.48 & 8.31 \\
M5 & No & 3.92 & -1.29 & 2.43 & 8.28 \\
M13 & Yes & 4.84 & -1.53 & 2.61 & 8.51 \\
NGC~6752 & Yes & 2.61 & -1.54 & 2.47 & 6.88 \\
NGC~2808 & ? & 7.91 & -1.14 & 3.25 & 8.24 \\
\enddata
\tablecomments{Table of cluster parameters of those investigated to host SCWD. 
$\log{(\rho_{\text{hmr}})}$ is the log of density inside the half-mass radius, 
and $\log{(\text{t}_{\text{r}_{\text{c}}})}$ is the log of core relaxation time ($\log{(\text{t}_{\text{r}_{\text{c}}})}$). 
Mass, [Fe/H] and $\log{(\text{t}_{\text{r}_{\text{c}}})}$ values are from \cite{Harris2010} and $\log{(\rho_{\text{hmr}})}$ values are from \cite{2018Baumgradt}.}
\label{Table: Various Cluster Parameters}
\end{deluxetable*}

In this analysis, we used deep and precise near-UV HST photometry and theoretical models to investigate for the presence of an excess of WD stars in the peculiar GGC NGC~2808.

We found an average excess of WDs 
of $\approx$ 60-65\%, when WDs are compared to MSTO stars, and depending on the assumed fraction of $He$-enhanced stars. The excess decreases to $\approx$ 30\% when comparing WDs to HB stars, and does not depend on the level of $He$ enhancement in this case.
By using the MSTO as reference, we found that on average the HB star excess is $\approx$ 45-50\%. In all three comparisons, we used a consistent set of theoretical models, i.e. isochrones, evolutionary tracks and WD cooling sequence from the BaSTI database. It is crucial to use models of the same input physics which therefore provide consistency and better control on systematics of the evolutionary time analysis.

The above ratios suggest a progressive deviation in number counts beyond the MSTO evolution in NGC~2808. 
Similar excesses in the number of HB and WD stars were found in the most massive GGC, $\omega$ Cen. \citet{2007Castellani} showed that HB stars are $\approx$ 40\% more abundant compared to MSTO stars and theoretical predictions, and this discrepancy decreases to 24-30\% when assuming a fraction of $He$-enhanced stars in the cluster. \citet{cal08mainpaper} found that the ratio of WD to MSTO star counts is at least a factor of 2 larger than the ratio of CO-core WD cooling to MS lifetimes. These results would suggest similarities between  $\omega$ Cen and NGC~2808, as previously proposed by \citet{johnson23}.




\cite{2021Chen} had studied the WD sequences in M~3 and M~13 using star count techniques and compared their results with BaSTI models. Their analysis suggested an excess of WDs in M~13 and not in M~3. The authors assigned this excess to the residual H envelope in stars with HB mass smaller than 0.56~M$_{\odot}$. These stars possibly skipped the AGB phase and so the third dredge-up, and maintained a residual thick H-envelope with sufficient mass \cite[of the order of 10$^{-4}$ M$_{\odot}$, see][]{2010Renedo} that sustained burning during the subsequent WD evolution, thus, leading to an increased cooling time (SCWDs). \cite{2021Chen} showed this leads to an observation of excess WDs. The authors related the existence of SCWDs to the existence of well-populated blue-HB (or an excess of EBT stars) in M~13, which is missing in M~3. They estimated that 25$\%$ of the HB consists of EBT stars. Following their previous works, \cite{2022Chen} carried out a similar analysis in NGC~6752 and predicted the existence of SCWDs in this cluster. The HB morphology of NGC~6752 and M~13 is similar, especially with the existence of blue-HB stars. The linkage between extended HBs and SCWDs was further validated by recent works of \cite{2023Chen}. For M~5, which has a similar HB morphology like M~3 with no EBT-like sequence, the authors found the WDs following the canonical sequence. 

{ NGC 2808 is observed to host EBT stars and empirically shares a similar HB morphology as M~13 \citep{2023KumarRanjan} and NGC~6752.}
This is measured by the R$_2$ parameter, i.e. the ratio of AGB stars to HB stars. As per \cite{2010Gratton}, R$_2$ correlates with HB-morphology. Smaller values of R$_2$ correspond to bluer extension of the HB. Adopting the total count of AGB stars (102 $\pm$ 10) from \cite{johnson23}, we estimated the R$_2$ parameter for NGC~2808 to be 0.076 $\pm$ 0.008\footnote{102/1344 = 0.076}, similar to those for NGC~6752 and M~13, 0.06-0.07 \citep[see references in][]{2022Chen}. Following the arguments proposed by the aforementioned studies, NGC~2808 makes a case for hosting SCWDs. Furthermore, the luminous tails of the EBT3 sequence for NGC~2808 could host post-HB stars in the AGB-manqu\'e phase (see extended Fig.~6 of \citealt{2021Chen}). These stars evolve off the HB phase by directly skipping the AGB phase and landing up on the WD sequence. A study of 18 post-HB stars by \cite{Prabhu_2021} in NGC~2808 based on UVIT and HST archival data compared to evolutionary tracks supports the hypothesis that most of these stars evolved from the EBT3 phase, with zero-age HB masses $<$ 0.50~M$_{\odot}$ (see their Fig.~14). Since these HBs skip the AGB phase, and hence any dredge-ups, it is possible that they would have retained a hot hydrogen envelope during the proto-WD stage and could contribute to the WD excess as SCWDs. Using the star counts presented in Table~\ref{Table HB}, we estimated the ratio of EBT3 stars to all the HB stars ($\beta$). 
    \begin{equation}
         \beta = \frac{\text{N}_{\text{EBT3}}}{\text{N}_{\text{RHB}} + \text{N}_{\text{EBT1}} + \text{N}_{\text{EBT2}} + \text{N}_{\text{EBT3}}}
     \end{equation}
For the internal field, $\beta_{\text{INT}}$ was found to be 11.5 $\pm$ 1\% and for the external field $\beta_{\text{EXT}}$ = 5.9. \cite{2021Chen} have estimated that 25\% of HB stars are EBT3 stars in M~13. 
Comparison of $\beta_{\text{INT}}$ of NGC~2808 with the $\beta$ estimated for M13 suggests that clusters with 
$\beta$ = 11 to 25$\%$ could possibly host SCWDs. 
Table~\ref{Table: Various Cluster Parameters} list the parameters (mass, metallicity, density and core relaxation time) of the GCs discussed in this section. We observed no clear trend in the commonly used cluster parameters among the cases that host SCWDs and the ones that do not host SCWDs, except in the similar HB morphology. However, this conclusion lacks statistics; not enough GCs with a wide range of parameters were explored.

It is also possible that the excess detected in our study in the very bright part of the WD cooling sequence is composed of a mix of SCWDs and a fraction of He-core WDs. \cite{2007Sandquist} studied the cumulative luminosity function of bright RGBs in NGC~2808 and found a deficiency of stars at the tip by comparing observations to models. As per the authors, these ``missing'' RGB stars may have undergone a massive mass loss and skipped the hot He-flash stage, thus, resulting in a degenerate He-core which would cool off as a He-core WDs. They suggested that if He-core WDs are indeed present in NGC~2808, they should be observable as an excess in the brighter part of the WD cooling sequence, since they cool slower than canonical CO-core WDs. Previous works on $\omega$ Cen \citep[see][]{2005Monelli, cal08mainpaper} found an excess of bright WDs in the cluster; additionally, HST UV CMDs in the $F275W$, $F336W$, and $F438W$ filters, showed a distinct redder WD cooling sequence (see Fig.~2 of \citealt{2013Bellini}) in addition to the canonical CO-core WD sequence. 
According to \cite{2013Bellini}, the redder sequence is dominated ($\approx$ 90 $\%$) by He-core WDs ($\approx$ 0.46~M$_{\odot}$ objects), possibly the progeny of an He-enhanced stellar population in the cluster,
while the bluer CO-core WD sequence ($\approx$ 0.55~M$_{\odot}$ objects) is the progeny of the He-normal population. They suggested that He-core WDs cool two times slower as compared to their CO-core counterparts. If He-core WDs are indeed present in NGC~2808, we would expect a similar redder WD cooling sequence as in $\omega$ Cen in the $F275W,\ F275W - F336W$ CMD. However, we did not detect such a sequence in this cluster based on the available HST UV data in the internal field, and nor in the external field. This could result from the lower precision of the HST photometric catalog of NGC~2808 since its distance is almost double as $\omega$ Cen distance. Future deeper HST UV data of NGC~2808 could shed light on this issue.
\section{Summary}\label{section:summary}


We used deep near-UV photometric data of NGC~2808 from the HUGS database, covering $\approx$ 1.5$\arcmin$ across the cluster center (the so-called internal field), and reduced WFC3-UVIS data for a field centered at $\approx$ 5$\arcmin$ away from the center (the so-called external field), to study the WD cooling sequence.
    
The $F275W,\ F275W - F336W$ CMD was used to identify candidate WDs, and these were verified to be stars and not artifacts by visually inspecting them on three-color image cut-outs (see Appendix~\ref{wd_cutouts}). 
We selected a sample of WDs in the external field by following the same method. Combining internal and external fields, we identified a total of 488 WDs, out of which 336 lie in the internal field and 152 lies in the external field.
    
We used BaSTI models (isochrones, evolutionary and WD cooling tracks), and employed a star count-evolutionary time analysis technique to estimate the excess or deficiency of observed WD stars compared to HB and MSTO stars. 
We found an excess of WDs compared to what predicted by models. Furthermore, by comparing HB to MSTO star count ratios to their evolutionary time ratios, we found that also HBs are $\approx$ 24-40\% in excess with respect to the MSTO, according to the fraction of $He$-enhanced stars considered.
The above ratios suggest a progressive deviation in number counts beyond the MSTO stars in NGC~2808. 
    
The comparison of our results with previous investigations in other GGCs, such as M~3, M~13, NGC~6752, and M~5, suggests that the excess could be possibly linked to the existence of SCWDs. The reason could be associated with the extended blue HB sequence observed in NGC~2808, similar to those observed in M~13 and NGC~6752. Such stars have HB mass smaller than 0.56~M$_{\odot}$, and could skip the third dredge-up to maintain a residual thick H-envelope that sustains burning during the subsequent WD evolution, thus, leading to increased cooling time. Our study suggests that NGC~2808 is the third GGC that could host SCWDs.

We did not detect any bimodality in color in the WD cooling sequence (from 21.4 to 24~mag) that might indicate the existence of He-core WDs as observed in $\omega$-Cen. Our analysis does not discard the existence of He-core WDs in NGC~2808 but suggests that if He-core WDs are present, they would be mixed with SCWDs. 
In future works, we will look into the RGB luminosity function (RGBLF), since, if He-core WDs are present in NGC~2808, the shape of the RGBLF would be different for NGC~2808 than a cluster which does not host He-core WDs. This would help us shed light on the percentage of stars in the WD cooling sequence which could be He-core WD. It would also help us understand if He-core WDs is a species only to be found in massive GGCs.

\begin{acknowledgements}
  { The authors thank the anonymous referee for their valuable comments on the manuscript that helped improve its quality.} The authors thank the $HST$ UV Globular Cluster Survey (HUGS) team for the photometric catalog of NGC~2808, Hubble Space Telescope Atlases of Cluster Kinematics (HACKS) team for proper motion data, and the Barbara A. Mukulski Archive for Space Telescopes (MAST, maintained by Space Telescope Science Institute, Baltimore, MD, USA) for making the above accessible. { The authors thank Maurizio Salaris (Astrophysics Research Institute, Liverpool John Moores University, Liverpool, UK) and Giuseppe Bono (Università di Roma Tor Vergata, Rome, Italy) for valuable comments and suggestions on the manuscript.}
\end{acknowledgements}

\FloatBarrier

\newpage

\appendix

\section{Differential Reddening} \label{reddening}

To check for differential reddening, we created an interpolated 2D extinction map (details below). The updated star counts were 1907 and 1945 in the LB and UB, respectively, without completeness correction. This is 3852 in total which is a $\approx$ 0.5\% change in total MSTO count without completeness correction (3872). This change does not affect the results of our analysis.

\begin{figure*}[ht!]
    \centering
    \includegraphics[width=0.7\linewidth]{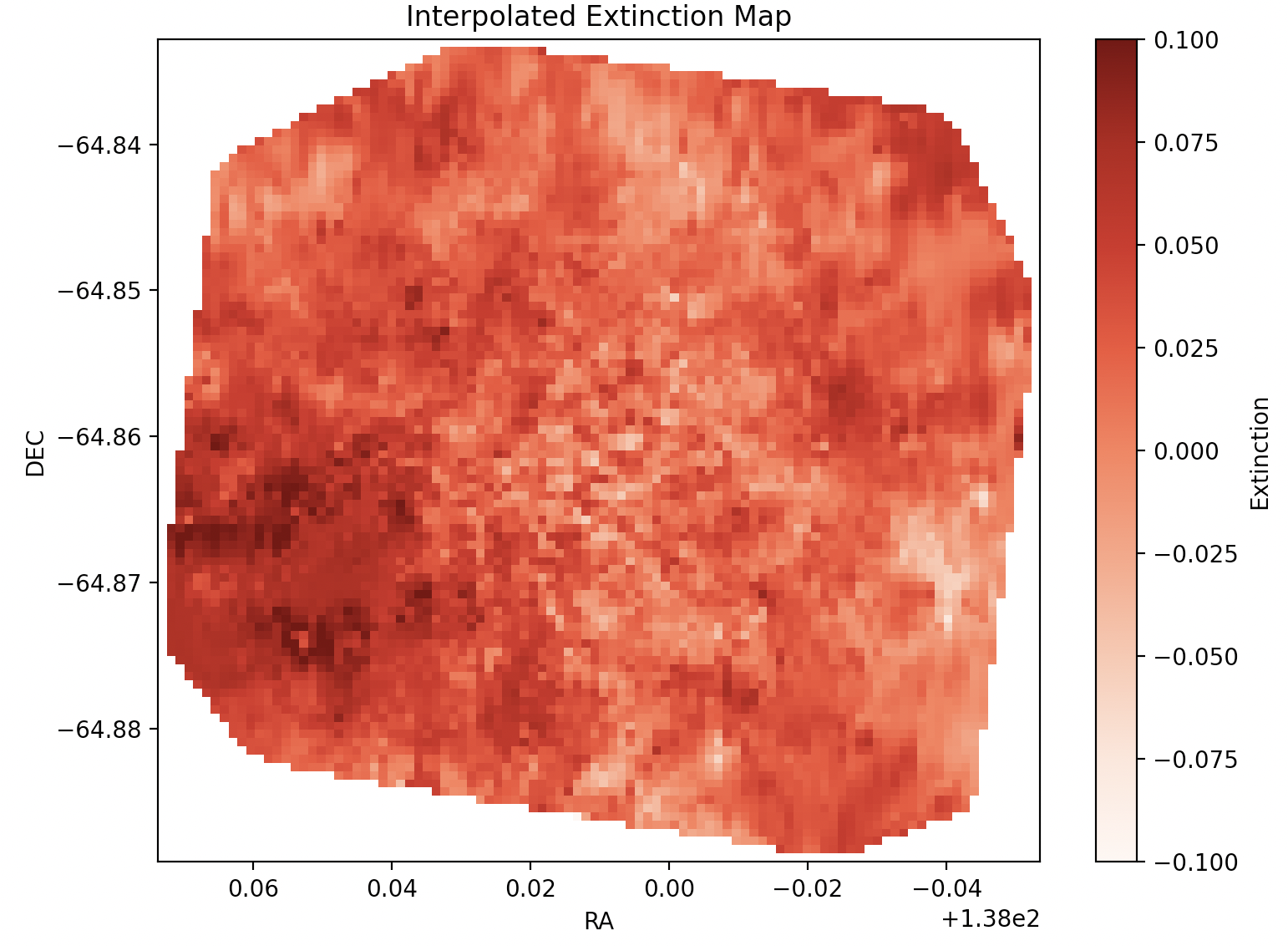}
    \caption{Differential Reddening Corrections}
    \label{DiffRed}
\end{figure*}

Although differential reddening in the line-of-sight towards NGC 2808 is generally small ($<$ 0.02 mag; \citep{bedin2000}), improvements can still be made to the photometry by correcting for localized reddening variations.  In order to build a differential reddening map for NGC~2808, we isolated the RGB sequence in the i versus g$-$i plane from Figure 4 of \citet{johnson23} and partitioned the data into 0.1 magnitude i-band bins. The median color in each i-band bin was calculated, and a fiduciary spline function was fit through the bins to serve as the reference median reddening sequence. Using the selective extinction coefficients from Vivas et al. (2017; see also \citep{SchlafyFinkbeiner}) and a median E(B-V) = 0.19 (\citep{bedin2000}), each star was slid along the reddening vector until it intersected the fiduciary sequence. The direction and distance required to move each star to the fiduciary sequence gave a measurement of delta E(B-V) along each sight line. The delta E(B-V) value for each star was then averaged with up to its 30 nearest neighbors within a 2 arc minute radius, and these mean values were then binned and interpolated to produce the smooth differential extinction map shown in Fig.~\ref{DiffRed}.


\section{WD color cut-outs images}\label{wd_cutouts}

\begin{figure*}[ht!]
\vspace{-0.5cm}
\begin{minipage}{0.24\textwidth}
\includegraphics[height=0.2\textheight,width=1.15\textwidth]{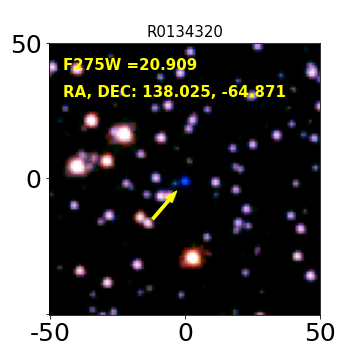} 
\end{minipage}
\begin{minipage}{0.24\textwidth}
\includegraphics[height=0.2\textheight,width=1.15\textwidth]{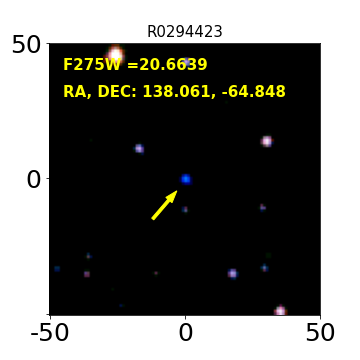} 
\end{minipage}
\begin{minipage}{0.24\textwidth}
\includegraphics[height=0.2\textheight,width=1.15\textwidth]{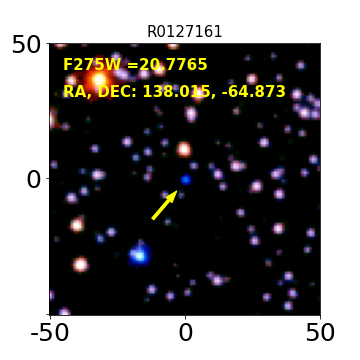} 
\end{minipage}
\begin{minipage}{0.24\textwidth}
\includegraphics[height=0.2\textheight,width=1.15\textwidth]{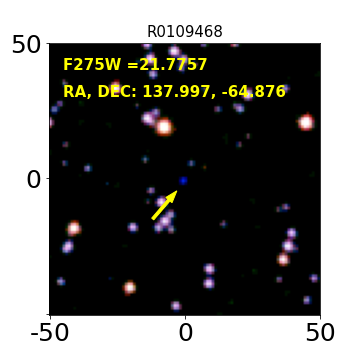} 
\end{minipage}
\begin{minipage}{0.24\textwidth}
\includegraphics[height=0.2\textheight,width=1.15\textwidth]{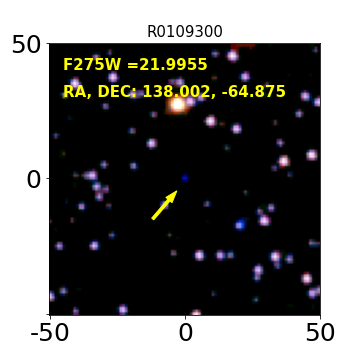} 
\end{minipage}
\begin{minipage}{0.24\textwidth}
\includegraphics[height=0.2\textheight,width=1.15\textwidth]{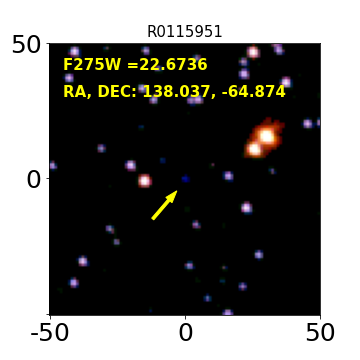} 
\end{minipage}
\begin{minipage}{0.24\textwidth}
\includegraphics[height=0.2\textheight,width=1.15\textwidth]{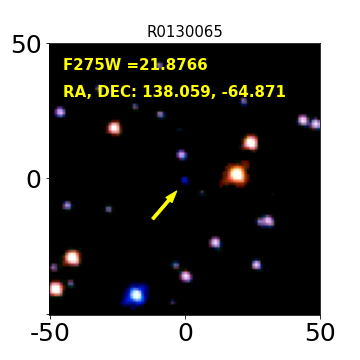} 
\end{minipage}
\begin{minipage}{0.24\textwidth}
\includegraphics[height=0.2\textheight,width=1.15\textwidth]{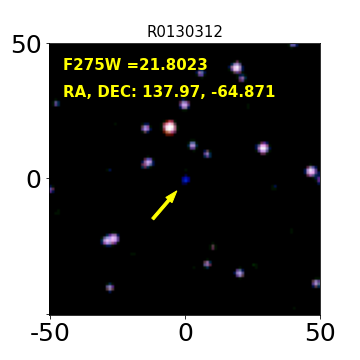} 
\end{minipage}
\begin{minipage}{0.24\textwidth}
\includegraphics[height=0.2\textheight,width=1.15\textwidth]{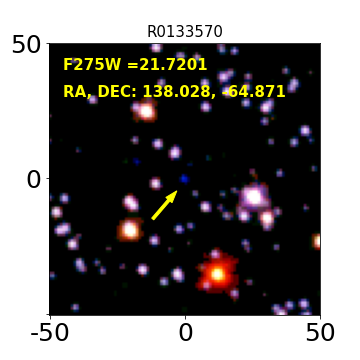} 
\end{minipage}
\begin{minipage}{0.24\textwidth}
\includegraphics[height=0.2\textheight,width=1.15\textwidth]{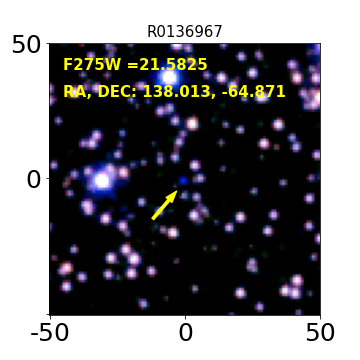} 
\end{minipage}
\begin{minipage}{0.24\textwidth}
\includegraphics[height=0.2\textheight,width=1.15\textwidth]{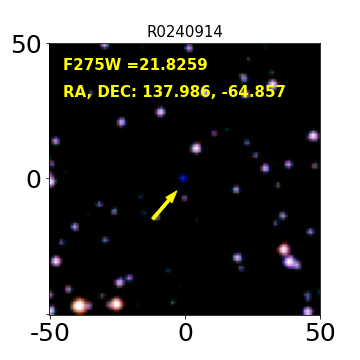} 
\end{minipage}
\begin{minipage}{0.24\textwidth}
\includegraphics[height=0.2\textheight,width=1.15\textwidth]{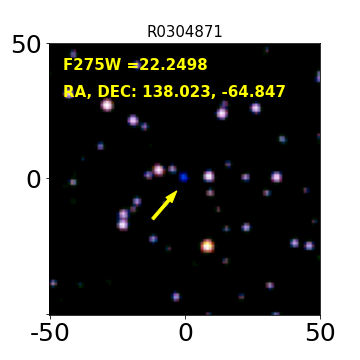} 
\end{minipage}
\begin{minipage}{0.24\textwidth}
\includegraphics[height=0.2\textheight,width=1.15\textwidth]{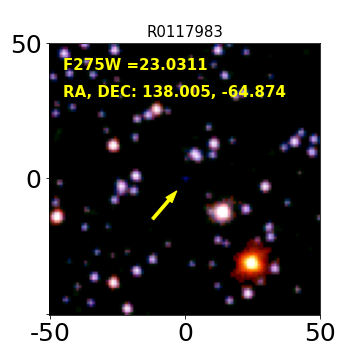} 
\end{minipage}
\begin{minipage}{0.24\textwidth}
\includegraphics[height=0.2\textheight,width=1.15\textwidth]{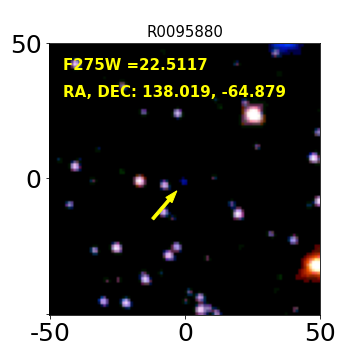} 
\end{minipage}
\begin{minipage}{0.24\textwidth}
\includegraphics[height=0.2\textheight,width=1.15\textwidth]{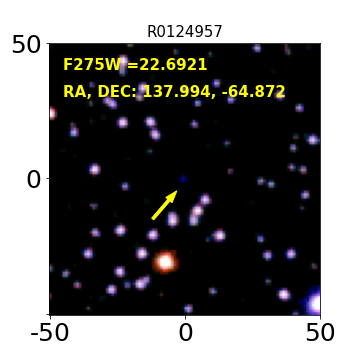} 
\end{minipage}
\begin{minipage}{0.24\textwidth}
\includegraphics[height=0.2\textheight,width=1.15\textwidth]{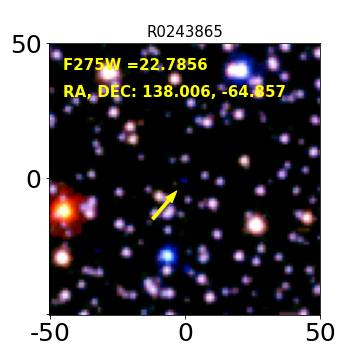} 
\end{minipage}
\begin{minipage}{0.24\textwidth}
\includegraphics[height=0.2\textheight,width=1.15\textwidth]{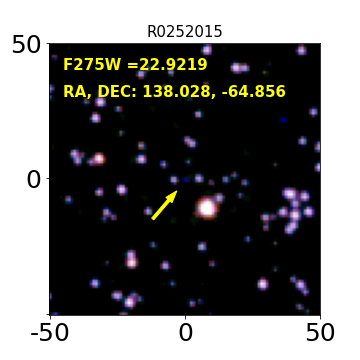} 
\end{minipage}
\hspace{0.07cm}
\begin{minipage}{0.24\textwidth}
\includegraphics[height=0.2\textheight,width=1.15\textwidth]{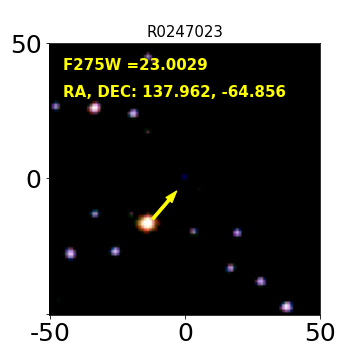} 
\end{minipage}
\hspace{0.07cm}
\begin{minipage}{0.24\textwidth}
\includegraphics[height=0.2\textheight,width=1.15\textwidth]{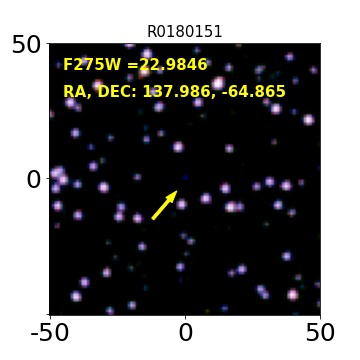} 
\end{minipage}
\hspace{0.07cm}
\begin{minipage}{0.24\textwidth}
\includegraphics[height=0.2\textheight,width=1.15\textwidth]{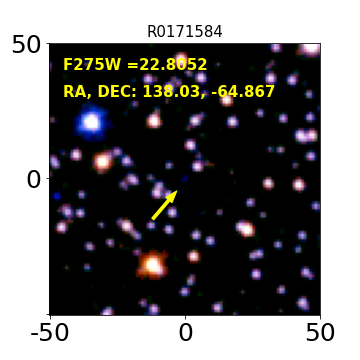} 
\end{minipage}
\caption{\scriptsize Cut-outs for 20 of the WDs identified in NGC~2808 based on a color image created from WFC3 drizzled images in the $F275W$, $F336W$, and $F438W$ filters.
Star $F275W$ magnitude and coordinates are labeled on the image. 
The yellow solid arrow points at the WD.  \label{fig:wd_rgb}}
\end{figure*}

We used the drizzled images of the internal field in the $F275W$, $F336W$, and $F438W$ filters provided in the HUGS archive to create a three-color stacked image. A region of 50$\times$50 pixels ($\approx$ 2$\times$2 arcsec) was selected around the position of each WD identified in the internal field and a postage stamp image was created; twenty of these are shown in Fig.~\ref{fig:wd_rgb}.
The WDs are at the center of the images and appear as very faint blue objects compared to the other stars in the cluster.


\section{Star count-crossing time analysis using various $\delta\mathcal{M}$ values}\label{section:deltaalphavalues}

As mentioned in Section~\ref{MSTO Analysis}, various values of $\delta\mathcal{M}$ were calculated. They are listed in \ref{tab:MSTO numbers with various delta alpha values}. Using these, WD-MSTO ratios were calculated as given in \ref{tab:WD MSTO ratios using various delta alpha values}. 

\FloatBarrier

\begin{deluxetable}{cccc}
\label{tab:MSTO numbers with various delta alpha values}
\tabletypesize{\scriptsize}
\tablecaption{Star counts and crossing times of MSTO stars for various $\delta\mathcal{M}$ values.}
\tablehead{
\colhead{$\delta\mathcal{M}$} & \colhead{Field} & \colhead{Count} & \colhead{Crossing Time (Myrs)}
}
\startdata
\multirow{2}{*}{0.0010} & INT & 1041 $\pm$ 32 & \multirow{2}{*}{191 $\pm$ 19} \\
 & EXT & 58 $\pm$ 8 & \\
\multirow{2}{*}{0.0020} & INT & 2087 $\pm$ 46 & \multirow{2}{*}{356 $\pm$ 36} \\
 & EXT & 113 $\pm$ 11 & \\
\multirow{2}{*}{0.0030} & INT & 3058 $\pm$ 55 & \multirow{2}{*}{522 $\pm$ 52} \\
 & EXT & 156 $\pm$ 12 & \\
\multirow{2}{*}{0.0035} & INT & 3541 $\pm$ 60 & \multirow{2}{*}{633 $\pm$ 63} \\
 & EXT & 188 $\pm$ 14 & \\
\multirow{2}{*}{0.0040} & INT & 4098 $\pm$ 64 & \multirow{2}{*}{687 $\pm$ 69} \\
 & EXT & 208 $\pm$ 14 & \\
\multirow{2}{*}{0.0050} & INT & 5065 $\pm$ 71 & \multirow{2}{*}{825 $\pm$ 83} \\
 & EXT & 255 $\pm$ 16 & \\
\multirow{2}{*}{0.0060} & INT & 6169 $\pm$ 79 & \multirow{2}{*}{1075 $\pm$ 107} \\
 & EXT & 310 $\pm$ 18 & \\
\enddata
\end{deluxetable}


\vspace{-0.25cm}

\begin{deluxetable}{cccccc}
\tabletypesize{\scriptsize}
\label{tab:WD MSTO ratios using various delta alpha values}
\tablecaption{WD-MSTO star count and crossing time ratios using various $\delta\mathcal{M}$ values. 
}
\tablehead{
\colhead{$\delta\mathcal{M}$} & \colhead{Field} & \colhead{Bin} & \colhead{$N_{\text{WD}}/N_{\text{MSTO}}$} & \colhead{$t_{\text{WD}}/t_{\text{MSTO}}$} & \colhead{Excess WD \%}
}
\startdata
\multirow{4}{*}{0.0010} & INT & Bright & 0.219 $\pm$ 0.016 & 0.067 $\pm$ 0.010 & 69.3 $\pm$ 4.9 \\
 & INT & Faint & 0.470 $\pm$ 0.026 & 0.148 $\pm$ 0.021 & 68.6 $\pm$ 4.8 \\
 & EXT & Bright & 0.747 $\pm$ 0.150 & 0.270 $\pm$ 0.038 & 63.8 $\pm$ 8.9 \\
 & EXT & Faint & 3.836 $\pm$ 0.566 & 0.867 $\pm$ 0.123 & 77.4 $\pm$ 4.6 \\
\multirow{4}{*}{0.0020} & INT & Bright & 0.109 $\pm$ 0.008 & 0.036 $\pm$ 0.005 & 67.0 $\pm$ 5.2 \\
 & INT & Faint & 0.234 $\pm$ 0.012 & 0.079 $\pm$ 0.011 & 66.2 $\pm$ 5.1 \\
 & EXT & Bright & 0.383 $\pm$ 0.068 & 0.145 $\pm$ 0.020 & 62.2 $\pm$ 8.6 \\
 & EXT & Faint & 1.969 $\pm$ 0.227 & 0.464 $\pm$ 0.066 & 76.4 $\pm$ 4.3 \\
\multirow{4}{*}{0.0020} & INT & Bright & 0.074 $\pm$ 0.005 & 0.025 $\pm$ 0.003 & 67.0 $\pm$ 5.2 \\
 & INT & Faint & 0.160 $\pm$ 0.008 & 0.054 $\pm$ 0.008 & 66.2 $\pm$ 5.1 \\
 & EXT & Bright & 0.278 $\pm$ 0.048 & 0.099 $\pm$ 0.014 & 64.4 $\pm$ 7.9 \\
 & EXT & Faint & 1.426 $\pm$ 0.149 & 0.317 $\pm$ 0.045 & 77.8 $\pm$ 3.9 \\
\multirow{4}{*}{0.0035} & INT & Bright & 0.064 $\pm$ 0.004 & 0.020 $\pm$ 0.003 & 68.5 $\pm$ 4.9 \\
 & INT & Faint & 0.138 $\pm$ 0.007 & 0.045 $\pm$ 0.006 & 67.7 $\pm$ 4.8 \\
 & EXT & Bright & 0.230 $\pm$ 0.039 & 0.082 $\pm$ 0.012 & 64.6 $\pm$ 7.8 \\
 & EXT & Faint & 1.184 $\pm$ 0.117 & 0.261 $\pm$ 0.037 & 77.9 $\pm$ 3.8 \\
\multirow{4}{*}{0.0040} & INT & Bright & 0.056 $\pm$ 0.004 & 0.019 $\pm$ 0.003 & 66.4 $\pm$ 5.3 \\
 & INT & Faint & 0.119 $\pm$ 0.006 & 0.041 $\pm$ 0.006 & 65.6 $\pm$ 5.1 \\
 & EXT & Bright & 0.209 $\pm$ 0.035 & 0.075 $\pm$ 0.011 & 64.0 $\pm$ 7.9 \\
 & EXT & Faint & 1.071 $\pm$ 0.103 & 0.241 $\pm$ 0.034 & 77.5 $\pm$ 3.8 \\
\multirow{4}{*}{0.0050} & INT & Bright & 0.045 $\pm$ 0.003 & 0.016 $\pm$ 0.002 & 65.5 $\pm$ 5.4 \\
 & INT & Faint & 0.097 $\pm$ 0.005 & 0.034 $\pm$ 0.005 & 64.6 $\pm$ 5.3 \\
 & EXT & Bright & 0.170 $\pm$ 0.028 & 0.063 $\pm$ 0.009 & 63.2 $\pm$ 8.0 \\
 & EXT & Faint & 0.873 $\pm$ 0.080 & 0.200 $\pm$ 0.028 & 77.0 $\pm$ 3.9 \\
\multirow{4}{*}{0.0060} & INT & Bright & 0.037 $\pm$ 0.002 & 0.012 $\pm$ 0.002 & 67.7 $\pm$ 5.1 \\
 & INT & Faint & 0.079 $\pm$ 0.004 & 0.026 $\pm$ 0.004 & 66.9 $\pm$ 4.9 \\
 & EXT & Bright & 0.140 $\pm$ 0.023 & 0.048 $\pm$ 0.007 & 65.6 $\pm$ 7.4 \\
 & EXT & Faint & 0.718 $\pm$ 0.063 & 0.154 $\pm$ 0.022 & 78.6 $\pm$ 3.6 \\
\enddata
\end{deluxetable}

\bibliography{Bibliography}{}
\bibliographystyle{aasjournalv7}

\end{document}